\documentclass[12pt]{article}
\usepackage{latexsym}
\usepackage{epsfig,amssymb,euscript}
\usepackage{amsmath}
\usepackage{array,calc,epsfig}
\usepackage[linktocpage]{hyperref}
\usepackage{pstricks}
\usepackage{bbm}
\usepackage{fancybox}

\def\be{\begin{equation}}
\def\ee{\end{equation}}

\def\bea{\begin{eqnarray}}
\def\eea{\end{eqnarray}}
\newcommand\bbone{\ensuremath{\mathbbm{1}}}

\numberwithin{equation}{section} 
\usepackage[]{graphicx}

\def\d {{\rm d}}

\def\calc         {{\cal C}}
\def\cald         {{\cal D}}

\def\calf         {{\cal F}}
\def\calg         {{\cal G}}

\def\calh         {{\cal H}}
\def\cali         {{\cal I}}
\def\calj         {{\cal J}}
\def\calk         {{\cal K}}
\def\call         {{\cal L}}
\def\calm         {{\cal M}}
\def\caln         {{\cal N}}
\def\calo         {{\cal O}}

\def\calr         {{\cal R}}

\def\calt         {{\cal T}}

\def\calw         {{\cal W}}
\def\calz         {{\cal Z}}

\def\1             {{\mathit 1}}
\def\2             {{\mathit 2}}

\def\H              {{\rm H}}

\def\del          {\partial}
\def\delbar       {\bar\partial}

\def\Re           {{\rm Re\hskip0.1em}}
\def\Im           {{\rm Im\hskip0.1em}}

\def\sqr#1#2{{\vcenter{\vbox{\hrule height.#2pt
 \hbox{\vrule width.#2pt height#1pt \kern#1pt \vrule width.#2pt}\hrule
 height.#2pt}}}}

\def\d{\text{d}}

\def\slashchar#1{\setbox0=\hbox{$#1$}           
\dimen0=\wd0                                 
\setbox1=\hbox{/} \dimen1=\wd1               
\ifdim\dimen0>\dimen1                        
\rlap{\hbox to \dimen0{\hfil/\hfil}}      
#1                                        
\else                                        
\rlap{\hbox to \dimen1{\hfil$#1$\hfil}}   
/                                         
\fi}

\begin{document}
\font\cmss=cmss10 \font\cmsss=cmss10 at 7pt

\vskip -0.5cm
\rightline{\small{\tt LMU-ASC 13/09}}

\vskip .7 cm

\hfill
\vspace{18pt}
\begin{center}
{\Large \textbf{ 
On moduli and  effective theory \\ of $N=1$ warped flux compactifications}}
\end{center}

\vspace{6pt}
\begin{center}
{\large\textsl{ Luca Martucci}}

\vspace{25pt}
\textit{\small Arnold Sommerfeld Center for Theoretical Physics,\\ LMU M\"unchen,
Theresienstra\ss e 37, D-80333 M\"unchen, Germany}\\  \vspace{6pt}
\end{center}

\vspace{12pt}

\begin{center}
\textbf{Abstract}

\end{center}

\vspace{4pt} {\small

\noindent  The moduli space of $N=1$ type II warped compactions  to flat space  with generic internal fluxes is studied. Using the underlying integrable generalized complex structure that characterizes these vacua, the different deformations are classified by $H$-twisted generalized cohomologies and identified with chiral and linear multiplets of the effective four-dimensional theory. The K\"ahler potential for chiral fields corresponding to classically flat moduli is discussed. As an application of the general results, type IIB warped Calabi-Yau compactifications and other SU(3)-structure subcases are considered in more detail.

}

\vspace{1cm}


\thispagestyle{empty}

\vfill
\vskip 5.mm
\hrule width 5.cm
\vskip 2.mm
{\small
\noindent e-mail: luca.martucci@physik.uni-muenchen.de
}

\newpage

\setcounter{footnote}{0}

\tableofcontents




\section{Introduction and summary}

In the context of string theory compactifications, a lot of recent attention has been focused on the study of the so-called flux compactifications  where the internal space contains fluxes in addition to branes and orientifolds,  see \cite{granareview,kachru,luest} for recent reviews. As a consequence, in supersymmetric flux compactifications the internal space is generically not Calabi-Yau (CY) and this complicates the identification of a low-energy effective theory.

In this context, a common approach in constructing  the effective theories considers fluxes (ordinary, geometric or non-geometric) and branes as ingredients added on top of an underlying CY-like geometry that determines the low-energy spectrum. The effect of fluxes then shows up through the appearance of a non-trivial potential and other new interactions giving mass to the light fields. This approach can be justified generically when these masses are small compared to the Kaluza-Klein scale of the underlying CY and this assumption can be rephrased  in the requirement that the backreaction of fluxes and branes should be negligible. This requirement is also generically needed in order to justify the constant  warp-factor approximation, which is usually assumed as well.\footnote{Reductions \`a la Scherk-Schwarz \cite{ss} are also possible in presence of enough symmetry of the internal space, although they are generically consistent truncations rather then low-energy effective actions.}  Clearly, even if the traditional approach can give perfectly trustable results, it may be too restrictive and exclude physically interesting settings, e.g.\ characterized by a non-trivial warping as discussed for example in \cite{gkp,giddings}.

 In this paper I would like to suggest an alternative strategy to study the low-energy effective theory of supersymmetric warped flux compactifications to flat space which is more similar in the spirit to the traditional approach to purely CY vacua. In ordinary CY compactifications, the low-energy spectrum is associated with the classical moduli describing the deformations of the  ten-dimensional supergravity preserving the CY condition. Although, these deformations are  usually described microscopically by harmonic forms, a lot of information on the low-energy theory depends on purely topological quantities, where  the harmonic forms describing the moduli can be safely substituted by any other representative in their cohomology classes. This feature clearly provides a great advantage, both practical and conceptual, and ultimately originates from the supersymmetry itself of the compactification, which implies that the space is CY, i.e.\ K\"ahler (which in turn implies complex and symplectic) with trivial canonical bundle.

Thus, as a first step in order to extend this approach to warped flux compactifications one would need to identify the available integrable structures dictated by the preserved supersymmetry, analogous to the complex and K\"ahler structures of ordinary CY's in the fluxless case, which could provide an organizing framework in which to study the moduli of flux compactifications. In particular, a desirable feature would be the existence of an associated cohomology (similar to Dolbeault cohomology for CY spaces) that could allow the identification of the moduli with different cohomology classes. 

Indeed, as shown in \cite{gmpt}, such a structure always exists for $N=1$ type II  compactifications to flat space with SU(3)$\times$SU(3) structure group and it coincides with a {\em generalized CY} structure, as defined by Hitchin in \cite{hitchin}. The existence of a generalized CY structure implies the existence of a {\em generalized complex} structure, whose definition extends and unifies the definitions of ordinary complex and symplectic structures. This in turn allows one to define a generalized Dolbeault operator with an associated cohomology. (See e.g.\ \cite{gualtieri,cavalcanti} for a detailed discussion of these aspects.)  These nice properties, although obviously intriguing, have not been concretely used so far in describing the low-energy supergravity theory of these general flux compactifications, at least not to my knowledge. 
More results in this direction are available if one restricts to the open string sector, whose 4D massless chiral spectrum has been identified in terms of particular cohomology classes defined by the background generalized complex structure \cite{lucapaul1}, and in  this paper I would like to provide evidence that something analogous happens for the closed string sector.  

In order to have a better idea of the nature of the cohomology classes arising in this context,  let us recall that one of the distinguishing features in this framework  is the use of polyforms, instead of forms of definite degree, as elementary objects. Then, in the presence of a non-trivial Neveu-Schwarz (NS) $H$-field, the natural differential acting on  polyforms is given by the $H$-twisted exterior derivative
\bea\label{dH}
\d_H:=\d +H\wedge\quad.
\eea  
Furthermore, the background generalized complex structure allows one to split $\d_H$ as $\del_H+\delbar_H$, where $\delbar_H$ is the generalized Dolbeault operator.  As we will see, the low-energy spectrum of the effective theory will be naturally described in terms of the twisted  cohomology classes of $\d_H$ and $\delbar_H$. This is compatible with what results from the study of topological sigma models with non-trivial $H$-field \cite{kapustin}, analogously to what happens in comparing the spectrum of physical D-branes in flux vacua \cite{lucapaul1} with the BRST spectrum of topological generalized complex branes \cite{kapustinli}.  

Unfortunately, a generalized CY structure contains, roughly speaking, only half of the structure available in ordinary CY spaces and thus is much weaker. Furthermore, its potential implications in the context of flux compactifications are still to be properly developed (see \cite{toma} for previous work in this direction\footnote{See also \cite{witt,cinese} for related discussions in the constant warping approximation.} and also as a useful introduction to this problem) and, on top of it, flux vacua with compact internal space generically require the presence of orientifolds, which complicate even more the already complicated mathematical problem. This will lead us to face a number of mathematical subtleties, like for example the proof of the validity of the so-called $\d\d^\calj$-lemma (see appendix \ref{lemma}) or the proof of the non-degeneracy of some extremization problems (see appendix \ref{fluxhitchin}). Although we will not provide a definitive answer to these sophisticated mathematical problems, we will adopt a pragmatical and physically motivated approach to circumvent them: the existence of a well-defined $N=1$ low-energy effective theory. This will naturally constrain the allowed possibilities, suggesting what should be the answer to these `microscopical' questions. Indeed, the interplay between ten-dimensional geometrical methods and four-dimensional effective description will play a crucial role in the following discussions and will eventually lead to a rather  unique and unambiguous picture.   
 
In this paper I will focus on closed string deformations, which can be encoded in two complex polyforms $\calz$ and $\calt$, where $\calz$ defines the generalized CY structure (and thus generalized complex structure) of the supersymmetric compactification. After reviewing in section \ref{sec:vacua} the structure of the supersymmetric vacua considered in this paper, in sections \ref{calz} and \ref{caltmoduli} we will see how a natural finite-dimensional parametrization of the $\calz$ and $\calt$ deformations is given by appropriate $H$-twisted  cohomology classes and will be associated with 4D chiral fields $z^I$ and $t^a$, respectively. In short, one can split the $\d_H$-cohomology into $\H^{\rm od}_H(M)\oplus \H_H^{\rm ev}(M)$. Then $z^I$ and $t^a$ parametrize 
\bea
\calm_\calz\simeq \H^{\rm od}_H(M;\mathbb{R})\quad\text{and}\quad\calm_{\calt}\simeq \H_H^{\rm ev}(M)\ ,
\eea
respectively.  An analogous characterization,  discussed in section \ref{linearsection}, is valid for the 4D linear multiplets $l_a$ dual to the chiral fields $t^a$. This identification will survive the check provided by the 4D effective coupling of different D-brane probes, which will be completely topological in nature and will exhibit the expected dependence on the closed string moduli  ---  see section \ref{probes}.  It turns out that $\calz$ (and thus the $z^I$ chiral fields) must generically satisfy additional restrictions derivable from a flux-generated  superpotential $\calw_{\rm eff}(z)$ and thus the corresponding physical moduli space reduces to 
\bea
\calm^{\rm flux}_\calz=\{z\in \calm_\calz:\d\calw_{\rm eff}(z)=0\}\ .
\eea On the other hand, in absence of  D-terms  generated by D-branes, the $\calt$-deformations described by $\calm_\calt$  turn out to be (classically) unobstructed moduli. So,  $\calz$ (or, more precisely the associated generalized complex structure) will be kept fixed, and  the attention will be restricted to the $\calt$-moduli. This simplifying assumption will guarantee the existence of a standard effective theory at sufficiently low energies and indeed it will be shown how an effective 4D (warped) K\"ahler potential for the $t^a$ chiral fields can be easily obtained by truncating the `microscopic' K\"ahler potential derived in \cite{lucapaul2}.\footnote{The K\"ahler potential of \cite{lucapaul2} can be seen as a warped version of the K\"ahler potentials  derived in  \cite{granalouis,grimm} in the same framework provided by generalized geometry but in the constant warping approximation. These papers and \cite{granareview} contain also a useful discussion about the  relation between these generalized K\"ahler potentials and other (unwarped) K\"ahler potentials obtained in the literature on flux compactification.} This will be discussed in section \ref{sec:kaehler}, where it will be shown how the resulting effective K\"ahler potential  satisfies some non-trivial consistency checks for which, remarkably, only the topological characterization of the moduli in terms of twisted cohomology classes will be important.  Further general aspects related to the effective K\"ahler potential, like the interpretation of the 4D no-scale condition in 10D terms  or the moduli-lifting effects generated by possible D-terms induced by D-branes, will be discussed in sections \ref{sec:noscale} and \ref{Dtermmoduli}.

Some subcases with SU(3)-structure will be considered somewhat more explicitly in sections \ref{example:wCY} and \ref{sec:othersu(3)}. In particular, section \ref{example:wCY} will be focused on the type IIB warped CY compactifications  \cite{gp,gubser,gkp}. In this case, by restricting $t^a$ to include the universal modulus and other moduli  corresponding to the $B$-field and the Ramond-Ramond (RR) $C_{\it 2}$, one can extract the explicit form of the corresponding  (warped) K\"ahler potential, which more generically is only implicitly defined   and appears to  depend on some microscopical details of the compactifications. The result is in agreement with the K\"ahler potential recently obtained in \cite{torroba}, and extends it to include also the $B$ and $C_{\it 2}$  moduli. Notice that the approaches followed here and in \cite{torroba} are completely different. The derivation of \cite{torroba} is based on a detailed dimensional reduction (along the lines  described in \cite{torroba0}) and does not use supersymmetry at all. On the other hand, in the derivation presented here supersymmetry plays a crucial role and allows the use of topological arguments which partly avoid the involvement of detailed microscopical conditions. 

The appendices \ref{twist}, \ref{purespinors} and \ref{lemma} summarize some background material about the framework used in this paper which could be useful for the non-expert reader before he starts reading section \ref{sec:vacua} (see also \cite{toma}).  In appendix \ref{app:orientifold} the effect of orientifolds, which is often considered implicit in the paper, is discussed in some detail and, finally, appendix \ref{fluxhitchin} discusses some Hitchin-like functionals which are extremized on (part of) the supersymmetry conditions.


\section{The structure of the $N=1$ vacua}
\label{sec:vacua}

In this section I briefly summarize the general properties of the type II $N=1$ vacua considered in this paper. The formalism used for describing these flux vacua, which adopts the language of generalized complex geometry, is essentially the one introduced in \cite{gmpt} but the conventions and definitions follow \cite{lucapaul2}. More details about these background aspects are given in appendix. See also appendix A of \cite{nonsusy} for a complete description of the supergravity conventions used here.

\subsection{The bosonic configuration}

We will consider the low-energy dynamics of general warped compactifications to flat four-dimensional space of type II theories. The ten-dimensional space has the structure $X_{10}=X_4\times M$ with coordinates $x^\mu$ and $y^m$ on $X_4$ and  $M$, respectively. The ten-dimensional metric splits as
\bea\label{metricsplit}
\d s^2_{X_{10}}=e^{2A}\d s^2_{X_4}+\d s^2_{M}\ ,
\eea
where the warp factor $A$ depends generically on $y^m$. We take as independent RR field-strengths only the internal ones, with all legs along $M$, and group them in a single polyform
\bea
F=\sum_{\mathit k}F_{\mathit k}
\eea 
with $0\leq k\leq 6$ even/odd in IIA/IIB.
They satisfy the Bianchi identity  
\bea\label{BI}
\d F=-j\ ,
\eea
where $j$ is the current associated with the different D-branes and orientifolds.  More explicitly, (in string units $2\pi\sqrt{\alpha^\prime}=1$) we have
\bea\label{DOcurrents}
j=\sum_{a\in\text{D-branes}}j^{\rm D}_a-\sum_{b\in\text{O-planes}}\tau_b j^{\rm O}_b\ ,
\eea 
with $\tau_{{\rm O}p}=2^{p-5}$. For a D-brane wrapping a cycle $\Sigma_a\subset M$ with U(1) field-strength F${}_a$ we have\footnote{In our convention the delta-function  is defined in terms of the Mukai pairing (see appendix \ref{twist}) by $\int_{M}\langle\omega,\delta(\Sigma)\rangle=\int_\Sigma\omega$, for any form $\omega$ of degree equal to the dimension of $\Sigma$.} 
\bea\label{Dcurrent}
j^{\rm D}_a=\delta(\Sigma_a)\wedge e^{-{\rm F}_a}\ .
\eea
Furthermore, in the presence of orientifolds, we consider $M$ as the covering space of the actual orientifolded space with fixed O-planes $O_b$ and associated currents $j^{\rm O}_b=\delta(O_b)$.  See appendix \ref{app:orientifold} for further details on the orientifold projections. Notice that in  the explicit form for the localized currents we are for simplicity omitting the higher order curvature corrections. They can be easily restated in all the expressions by replacing $j\rightarrow j^{\text{new}}=j\wedge (\text{curv.corr.})$.\footnote{For multiple coincident D-branes the currents should be further modified,  for example by replacing $e^{-{\rm F}_a}$ with the Chern character ${\rm ch}(-{\rm F}_a)$.} 

In this section and in most of the paper we use twisted polyforms  which transform as $(\ldots)\rightarrow e^{\d \lambda}\wedge(\ldots)$ under the gauge transformation $B\rightarrow B+\d\lambda$ ---  see appendix \ref{twist} for further details. The natural differential acting on them is just the usual exterior derivative $\d$. Occasionally, when explicitly stated, we will use other equivalent pictures described in appendix \ref{twist}, where the differential is twisted by the $H$-flux, as in (\ref{dH}).

The background RR field-strengths have external components given (in polyform notation) by  $\d{\rm vol}_{X_4}\wedge (e^{4A}*_B F)$, where $*_B$ is the twisted six-dimensional Hodge-star operator defined in (\ref{Bhodge}). Notice that the equations of motion require that 
\bea\label{RReq}
\d (e^{4A}*_B F)=0\ .
\eea

\subsection{Pure spinors, $N=1$ conditions and generalized complex structure}

A generic warped  flux compactification with 4D space-filling  D-branes and orientifolds can be completely characterized by two O(6,6) pure spinors $\calz$ and $T$ (not to be confused with $\calt$), defining an SU(3)$\times$SU(3)-structure. They are complex polyforms on $M$ of opposite parity
\bea
\calz=\sum_{{\mathit k}\text{ even/odd}}\calz_{\mathit k}\quad\quad,\quad\quad T=\sum_{{\mathit l}\text{ odd/even}}T_{\mathit l}\ ,
\eea
with ${\mathit k}$ even (odd) and  ${\mathit l}$ odd (even) in IIA (IIB)  ---  see appendix \ref{purespinors} for more details about them. In the twisted picture we are using here, $\calz$ and $T$  contain the complete information about the NS sector, i.e.\ internal metric, $B$-field, warping and dilaton, as well as information about the reduced SU(3)$\times$SU(3)-structure of the doubled spin structure of type II theories, which will eventually be constrained by the supersymmetry condition. 

Using these variables, the background supersymmetry conditions  for compactifications to four flat dimensions   \cite{gmpt}  can be divided into three parts.\footnote{In this paper we are focusing on very general backgrounds, where however the two internal spinors describing the residual supersymmetry in \cite{gmpt} are assumed to have the same norm. In physical terms, this is equivalent to requiring that these backgrounds admit the introduction of supersymmetric D-branes and orientifolds or, in other words, which they are characterized by D-brane generalized calibrations \cite{lucal}.} First, one needs to require that
\bea\label{susy1}
\d\calz=0\ .
\eea
This means that the internal space is an integrable generalized Calabi-Yau as defined in \cite{hitchin} and  in turn implies that the associated generalized complex structure $\calj$  is integrable (cf.~appendix \ref{purespinors}). 


The second condition can be written in the form \cite{toma}
\bea\label{susy2bis}
\d\Re T=-\calj\cdot F\ .
\eea
Here $\calj$ acts  on polyforms as briefly described in appendix \ref{purespinors} (more details can be found e.g.\ in \cite{toma}). Equivalently, using the decomposition (\ref{ghodge}) and the integrability of $\calj$, we can write (\ref{susy2bis}) as
\bea\label{susy2}
F_{-1}=-i\bar\partial\Re T\quad,\quad F_{-3}=0\ .
\eea
The remaining background condition is  
\bea\label{dflat}
\d(e^{2A}\Im T)=0\ .
\eea
Notice that the conditions (\ref{susy2}) and (\ref{dflat}) automatically imply (\ref{RReq}).

As discussed in \cite{lucapaul2} and briefly reviewed in the following sections, the conditions (\ref{susy1}) and (\ref{susy2bis}) [or equivalently (\ref{susy2})] have a direct four-dimensional interpretation  as F-flatness conditions while the condition (\ref{dflat})  can be interpreted as D-flatness condition associated with the RR-symmetry $C\rightarrow C+\d\lambda$, which is gauged in the four-dimensional theory (cf.~appendix \ref{fluxhitchin}). 

In order to preserve supersymmetry, D-branes and orientifolds must be calibrated \cite{paulk,lucal,pauldimi}. This condition can  in turn be split into two parts \cite{lucal} which can be interpreted as F-flatness and D-flatness \cite{lucasup} (see sections \ref{calz} and \ref{Dtermmoduli} below). The first, interpreted as an F-flatness condition, says that supersymmetric D-branes and orientifolds wrap generalized complex cycles as defined in \cite{gualtieri}, i.e.\
\bea\label{gcsources}
j\in U_0\quad\Leftrightarrow\quad \calj\cdot j=0\ .
\eea
Notice that, by using the integrability of $\calj$, this condition  also  follows directly from (\ref{BI}) and (\ref{susy2bis}) since they imply that 
\bea\label{tbianchi}
\d\d^\calj \Re T=j\ ,
\eea
where $\d^\calj$ is defined in (\ref{dj}). The second condition is
\bea\label{braneDterm}
\langle\Im T,j\rangle=0\ ,
\eea
and can be interpreted as a D-flatness condition.
 
It is important to stress that all the supergravity equations of motion are satifsfied once the above supersymmetry conditions and the RR Bianchi identity (\ref{BI}) are imposed \cite{luestdimi,gauntlett,pauldimi}. Furthermore, compact spaces will generically require orientifolds and thus  $M$ must be rather considered as the covering space of the  internal space. All the fields, polyforms and the corresponding cohomology classes must satisfy appropriate projection conditions that are discussed in detail in appendix \ref{app:orientifold}. In the following, in order not to overload the general discussion,   the orientifold projection will often be considered as implicit and will be explicitly mentioned only if necessary. In any case, the effect of O-planes can be easily taken into account by applying the rules of appendix \ref{app:orientifold} and a more explicit example of their effect is provided  in the subcases discussed in sections \ref{example:wCY} and \ref{sec:othersu(3)}. 

 

\section{$\calz$-moduli  and the superpotential}
\label{calz}

The conditions  (\ref{susy1}) and (\ref{susy2}) can be derived as F-flatness conditions \cite{lucapaul2} from the superpotential 
\bea\label{totsup}
\calw=\int_{M}\langle\calz, F+i\d \Re T\rangle\ .
\eea 
In the `microscopic' untruncated four-dimensional picture adopted in \cite{lucapaul2},\footnote{See \cite{granalouis,grimm} for previous work  based on the same philosophy.} one must consider as closed string chiral fields $\calz$ itself and
\bea
\calt:=\Re T-iC\ ,
\eea
where the RR gauge potential $C$ is identified by splitting $F=F^{0}+\d C$ for some reference $F^{0}$. Indeed, $\calz$ and $\calt$ contain the full information about the background configuration and their complex fluctuations are given by
\bea\label{gde}
\delta \calz\in U_3\oplus U_1\quad,\quad \delta\calt\in U_0\oplus U_{-2}\ .
\eea 
Imposing $\delta_{\calt_{0}}\calw=\delta_{\calt_{-2}}\calw=0$ one gets (\ref{susy1}), while $\delta_{\calz_{3}}\calw=\delta_{\calz_{1}}\calw=0$
give (\ref{susy2}).

The moduli space of the generalized CY structure defined by equation (\ref{susy1}) has been studied already in \cite{hitchin}.  Assuming the $\d\d^\calj$-lemma [cf.~appendix \ref{lemma}], one can prove \cite{hitchin} (see also \cite{toma} for a discussion in our context) that the space of solutions to (\ref{susy1}) can be locally identified with the $H$-twisted cohomology class
\bea\label{cmoduli}
\calm_{\calz}\simeq \H_H^{\rm od}(M;\mathbb{R}) \ .
\eea 
 Using (\ref{cohodec}), one can define the complex structure on $\calm_{\calz}$ by identifying, at any point $\calz\in\calm_{\calz}$, the $(1,0)$-tangent bundle  with  
$\H^3_H(M)\oplus \H^1_H(M)$. $\H^3_H(M)$ gives an overall constant rescaling of $\calz$, which corresponds to the conformal compensator in the four-dimensional superconformal effective theory, while $\H^1_H(M)$ describes the infinitesimal deformations of the generalized complex structure $\calj$ defined by $\calz$ \cite{gualtieri,li,toma}. This can be directly seen  by using the $\delbar$-cohomology and the fact that $\H^k_{\delbar}(M)\simeq \H^k_H(M)$.

Notice that the presence of cohomology classes on the right-hand side of (\ref{cmoduli}) takes into account the identification of configurations related by the action of the group of generalized diffeomorphisms  $\calg$. This can be defined as the group extension 
\bea\label{gendiff}
0\rightarrow(B\text{ gauge transf.})\rightarrow \calg\rightarrow {\rm Diff}_0(M)\rightarrow 0
\eea
and combines the ordinary diffeomorphisms with the $B$-field gauge transformation $B\rightarrow B+\d\lambda$, acting on  polyforms by wedge-product with $e^{\d\lambda}$.  Clearly, $\calg$ is an infinite-dimensional symmetry group of our equations and relates different solutions which should be considered as physically equivalent.  The infinitesimal deformation of $\calg$ acting on a twisted polyform $\omega$ is given by $\delta_\mathbb{X}\omega=\call_\mathbb{X}\omega:=\d(\mathbb{X}\cdot \omega)+\mathbb{X}\cdot(\d\omega)$, where  $\mathbb{X}\in \Gamma(E)$  is a generalized vector field (cf. appendix \ref{twist}). 

As discussed in \cite{hitchin}, $\calm_{\calz}$ has a natural special K\"ahler (ant thus complex) structure. We may parametrize $\calm_\calz$ as follows, along the lines of what is done for ordinary CY spaces (see e.g.\ \cite{candelas}). Let us first introduce a  basis $\alpha_I,\beta^J$ (with $I,J=0,\ldots,n$) for  $\H_H^{\rm od}(M;\mathbb{R})$  (of even real dimension $2n+2$) such that
\bea
\int_{M}\langle\alpha_I, \beta^J\rangle=\delta_I{}^J\ .
\eea
Furthermore, we may assume $\alpha_I$ and $\beta^J$ to be  integral, in the sense that 
\bea
\label{quantcond} \int_\Sigma\alpha_I|_\Sigma\wedge e^{\rm F}\in \mathbb{Z}\quad,\quad \int_\Sigma\beta^J|_\Sigma\wedge e^{\rm F}\in \mathbb{Z}\ ,
\eea
where $(\Sigma,{\rm F})$ is  any generalized cycle \cite{lucajarah}.\footnote{In (\ref{quantcond}) curvature corrections $\sqrt{\hat A(T_\Sigma)/\hat A(N_\Sigma)}$ have been omitted for notational simplicity. See comment below (\ref{Dcurrent}).} Then, one can expand
\bea\label{calzdec}
\calz=z^I\alpha_I-\calg_J\beta^J\ ,
\eea
where 
\bea
z^I=\int_M\langle \calz, \beta^I\rangle\quad,\quad \calg_J=\int_M\langle \calz, \alpha_J\rangle\ .
\eea
Since $\calm_\calz$ has complex dimension $n+1$, in analogy with what happens for ordinary CY spaces, it is natural to assume that $z^I$ are good local holomorphic coordinates for $\calm_\calz$,\footnote{The $z^I$'s can also be considered as projective coordinates for the moduli of the generalized complex structure $\calj$.} so that $\calg_J=\calg_J(z)$. Since $\partial_I\calz\in U_3\oplus U_1$, as in the CY case, we have
\bea
2\calg_I(z)=\partial_I(z^J\calg_J)
\eea
so that $\calg_I=\partial_I\calg(z)$ for a certain holomorphic prepotential $\calg(z)$ which is homogeneous of degree two and encodes the special K\"ahler structure of $\calm_\calz$. Notice that, although this parametrization depends only on the cohomology of $\calz$, the cohomology representatives of $\alpha_I$ and $\beta^J$ in (\ref{calzdec}) are actually fixed (up to the generalized diffeomorphisms (\ref{gendiff})) by the requirement that $\calz$ is an O(6,6) pure spinor \cite{hitchin}.

We can now go back to the superpotential (\ref{totsup}) and  try to integrate out the massive modes contained in $\calt$ by directly imposing (\ref{susy1}) on it. The resulting effective superpotential is
\bea\label{supeff}
\calw_{\rm eff}=\int_{M}\langle\calz, F\rangle\ .
\eea 
Notice that the superpotential (\ref{supeff}) also contains  information about D-branes. Indeed we can split
\bea\label{fsplit}
F=F^{\rm back}+\theta\ ,
\eea
where $\d F^{\rm back}=0$ and $\theta$ is the generalized current of the form (\ref{Dcurrent}) associated with a generalized chain \cite{lucajarah} whose boundary coincides with the sum (with appropriate signs) of the local sources, so that $\d\theta=-j$. Then one  can split 
\bea\label{supsplit}
\calw_{\rm eff}=\calw^{\rm back}_{\rm eff}+\calw_{\text{D-branes}}\ ,
\eea
where $\calw^{\rm back}_{\rm eff}$ has the same form as (\ref{supeff}) but with $F^{\rm back}$ instead of $F$, and we have isolated the D-brane superpotential 
\bea\label{Dsup}
\calw_{\text{D-branes}}=\int_{M}\langle\calz, \theta\rangle=-\int_{\Gamma}\calz|_{\Gamma}\wedge e^{\tilde{\rm F}}\ ,
\eea
where $(\Gamma,\tilde{\rm F})$ is the generalized chain associated with the current $\theta$, whose boundary contains the D-brane generalized cycles. $\calw_{\text{D-branes}}$ coincides with the superpotential derived in \cite{lucasup} directly from the D-brane effective action. By extremizing it with respect to the open string degrees of freedom, one gets (\ref{gcsources}). Notice that the split (\ref{fsplit}) has an intrinsic ambiguity under the simultaneous shift $F^{\rm back}\rightarrow F^{\rm back}+\chi$ and $\theta\rightarrow \theta-\chi$, where $\chi$ defines any integral class in $\H^{\rm ev}_H(M;\mathbb{R})$. This ambiguity leads to an ambiguity  in the separate definitions of $\calw^{\rm back}_{\rm eff}$ and $\calw_{\text{D-branes}}$ and only the full superpotential (\ref{supeff}) is unambiguously  defined.

For simplicity, in most of the following discussions we do not explicitly consider the D-brane contribution to the complete superpotential (\ref{supeff}) or, in other words, we assume that we can always keep $j\in U_0$. This condition is automatically satisfied if there are only O-planes as localized sources. Thus, (\ref{supeff}) reduces to a superpotential $\calw_{\rm eff}(z)$ for $n+1$ chiral fields $z^I$ of Weyl weight 3, which include the conformal compensator corresponding to their overall rescaling. Notice that $\calw_{\rm eff}(z)$ is a superpotential of a superconformal supergravity. Once the compensator is eliminated by going to the Einstein-frame (see e.g.\ \cite{toine}), this gives a usual Einstein-frame superpotential, which is a section of a line bundle over the moduli space of the generalized complex structure $\calj$.

 If for example we apply this formalism to a non-compact internal manifold with no localized sources, which is a somewhat limiting case, then 
\bea\label{effnc}
\calw_{\rm eff}(z)=M_Iz^I-N^J\calg_J(z)\ ,
\eea
where
\bea\label{rrquanta}
M_I=\int_M\langle \alpha_I,F \rangle\quad,\quad N^J=\int_M\langle \beta^J,F \rangle\ .
\eea
However, when the internal  space is compact and thus there are at least O-planes, the application of the explicit expression (\ref{effnc}) requires some caution, because the RR-flux $F$ does not straightforwardly identify a $\d_H$-cohomology class.\footnote{An expansion like (\ref{effnc}) can be safely applied to $\calw^{\rm back}_{\rm eff}$ in (\ref{supsplit}) but, as stressed above, only the complete $\calw_{\rm eff}$ is physically meaningful.} 

In any case, the extremization of the superpotential $\calw_{\rm eff}(z)$ potentially lifts all the $z^I$ moduli, up to their overall rescaling corresponding to the four-dimensional  compensator. This can be understood at the `microscopical' level as follows. First, $F_{-3}$ clearly represents a class in $\H^{-3}_{\delbar}(M)$. On the other hand, the $\del\delbar$-lemma implies that we can write $\partial F_{-3}=\del\delbar \beta_{-2}$ and then from $\d F=-j\in U_0$ we see that $F_{-1}-\partial \beta_{-2}$ represents a class in $\H^{-1}_{\delbar}(M)$. Now, by considering the infinitesimal deformations of $\calz$ as described by $\H^{3}_{\delbar}(M)$ and $\H^1_{\delbar}(M)$, the extremization of $\calw_{\rm eff}$ requires that $\calz$ must be `aligned' in such a way that the classes in  $\H^{-3}_{\delbar}(M)$ and $\H^{-1}_{\delbar}(M)$ defined by $F_{-3}$ and $F_{-1}$ are trivial.   This condition constrains $\calz$ (but not its overall normalization) to lie on a subset of $\calm_\calz$. It is indeed necessary  in order for the 10D conditions (\ref{susy2}) to admit a solution and it is natural to conjecture that, under reasonable assumptions, it is actually sufficient too.

To summarize, we arrive at the following flux-modified $\calz$-moduli  space:
\bea
\calm^{\text{flux}}_{\calz}=\{z\in \calm_\calz\text{ such that }\d\calw_{\rm eff}(z)=0\}\ .
\eea
In other words, the RR fluxes can in principle completely fix  the generalized complex structure $\calj$. In the following we will most of the time assume that it indeed happens, writing 
\bea\label{fixedz}
\calz=Y^3\calz^{0}\ ,
\eea
where $Y$ is the conformal compensator of Weyl weight 1, and $\calz^{0}$ is a fixed-reference pure spinor that does not transform under Weyl transformations. Notice that, having fixed $\calz$ up to an overall rescaling, the symmetry group (\ref{gendiff}) is broken. More explicitly,  the generic infinitesimal deformation of $\calz$ under (\ref{gendiff}) is $\delta\calz=\d(\mathbb{X}\cdot\calz)=\partial(\mathbb{X}\cdot\calz)+\bar\partial(\mathbb{X}\cdot\calz)$ with $\partial(\mathbb{X}\cdot\calz)\in U_3$ and $\bar\partial(\mathbb{X}\cdot\calz)\in U_1$. Then one must impose $\bar\partial(\mathbb{X}\cdot\calz)=0$ and thus also $\partial\delbar(\mathbb{X}\cdot\calz)=0$. But, using the  $\del\delbar$-lemma (\ref{forlemma2}) this means that in fact $\delta\calz=\d(\mathbb{X}\cdot\calz)=0$ and then the residual symmetry of (\ref{gendiff}) is the subgroup $\calg_{\calz}$ that leaves  $\calz$ completely unchanged:
\bea\label{resym}
\calg_\calz=\{g\in\calg : g(\calz)=0\}\ .
\eea
In the following sections we will see how minimal $N=1$ supersymmetry naturally requires that $H^{2}_{\delbar}(M)=0$. In this case, the residual symmetry is generated by generalized vector fields $\mathbb{X}$ such that $\mathbb{X}\cdot\calz=\d(f\calz)$ for some function $f$.

Notice that the effective potential (\ref{supeff}) does not exactly reproduce the Gukov-Vafa-Witten superpotential \cite{gukov}
\bea\label{gvw}
\calw_{\rm GVW}=\int_M\Omega_{\rm CY}\wedge(F_{\it 3}+i\,e^{-\Phi} H)\ .
\eea
in the subcase of warped IIB CY compactifications.\footnote{On the other hand, $\calw_{\rm GVW}$ can be obtained from (\ref{totsup}) by truncating it in the naive way. This is  clearer in the untwisted picture, where $\d$ is substituted with $\d_H$ (cf.~appendix \ref{twist}). Then, by replacing $\calz$ with the CY holomorphic $(3,0)$-form $\Omega_{\rm CY}$ in (\ref{totsup}), but without assuming the stronger condition $\d_H\calz\equiv H\wedge \Omega_{\rm CY}=0$, one gets exactly (\ref{gvw}).} The origin of this difference is that  we use $H$-twisted cohomologies,  which already incorporate part of the effects of the $H$-field. In particular, this makes the axion-dilaton disappear in the effective superpotential ---  see section \ref{example:wCY} for more comments on it.


\section{$\calt$-moduli and massless chiral multiplets}
\label{caltmoduli}

Let us now see how the generalized complex structure $\calj$ allows an easy characterization of the $\calt$-moduli. We will work at fixed $\calz$, up to an overall rescaling corresponding to the conformal compensator, as in (\ref{fixedz}). As discussed in section \ref{calz}, the $\calz$ moduli-space $\calm^{\rm flux}_\calz$  can be potentially reduced to a discrete set (up to the compensator) by the superpotential (\ref{supeff}), and thus the $\calt$-moduli space $\calm_\calt$ will actually give the complete physical closed string moduli space. Less generically, the full moduli space will be a fibration of $\calm_\calt$ over $\calm^{\rm flux}_\calz$. 

The polyform $\calt$ contains information about the RR-potential $C$ and the `stable' \cite{hitchin} polyform $\Re T$, which must satisfy (\ref{compa2}). The associated allowed closed string deformations are thus given by $\delta \Re T\in U_0$ and a generic RR-deformation $\delta C=\delta C_0+(\delta C_{-2}+\text{c.c.})$.

First of all, pure RR moduli are given by closed finite shifts $\Delta C$. Taking into account the local RR-gauge symmetry  $\Delta C\rightarrow \Delta C+\d\Lambda$, the physically inequivalent RR-shifts are identified by
\bea
[\Delta C]\in \H_H^{\rm ev}(M;\mathbb{R})\ ,
\eea
which must be further modded out by  $\H_H^{\rm ev}(M;\mathbb{Z})$ since $[\Delta C]\in \H_H^{\rm ev}(M;\mathbb{Z})$ is physically equivalent to the zero class.\footnote{Here possible torsion contributions to $\H_H^{\rm ev}(M;\mathbb{Z})$ are ignored.} Thus, the RR-shifts parametrize a torus
\bea\label{rrm}
\calm_{\rm RR}\simeq\H_H^{\rm ev}(M;\mathbb{R})/\mathbb{Z}^{b^{\rm ev}}
\eea
of dimension $b^{\rm ev}:=\dim \H_H^{\rm ev}(M;\mathbb{R})$.

Notice that, using the $\d\d^\calj$-lemma we can split [cf.~appendix \ref{lemma}]
\bea
\H_H^{\rm ev}(M)=\H_H^2(M)\oplus \H_H^0(M)\oplus \H_H^{-2}(M)\ .
\eea
In particular, we can parametrize the RR moduli just in terms of $\H_H^{-2}(M)$ and the real elements in $\H_H^0(M)$. However, $\H^{-2}_H(M)$ has a quite different nature from $\H^{0}_H(M)$ since the RR-shifts in $\H^{-2}_H(M)$ do not naturally combine with the NS degrees of freedom contained in $\Re T\in U_0$ to give 4D chiral fields. As we will recall in section \ref{sec:kaehler}, the $N=1$ K\"ahler potential for our vacua can depend in a direct way only on the NS degrees of freedom \cite{lucapaul2} and thus the chiral fields associated with the possible moduli $\H^{-2}_H(M)$ would  not find a natural 4D interpretation in a strictly minimal (i.e.\ $N=1$) supersymmetric setting. One way to understand this from a microscopical point of view is to notice that $\H^{-2}_H(M)$ originates from fluctuations that transform in the ${\bf 3}$ or ${\bf \bar 3}$ representation of the SU(3)$\times$SU(3) structure group underlying the $N=1$ compactification and are thus  not `natural' if supersymmetry is minimal.\footnote{See \cite{granalouis,grimm} for analogous arguments in the untruncated formulation of those papers.} This is also consistent with experience from ordinary CY orientifold compactifications \cite{louis}, where RR moduli are always completed into 4D chiral fields by NS moduli. These observations suggest that $N=1$ supersymmetry implies that
\bea\label{cohocond}
\H^{-2}_H(M)={0}\quad[\Leftrightarrow\quad \H^{2}_H(M)={0}]\ .
\eea
This property is analogous to the well known fact that $h^{2,0}=h^{3,1}=0$ for ordinary  CY spaces (with strict SU(3)-holonomy).\footnote{Something similar to (\ref{cohocond}) happens  in the somewhat different context of flux compactifications to AdS$_4$ spaces (which have no integrable generalized complex structure) studied in \cite{amir,paul,davide}, where the truncation on nilmanifolds and coset spaces is considered.}  In the following, (\ref{cohocond}) will always be assumed  to hold. See sections \ref{example:wCY} and \ref{sec:othersu(3)} for  additional discussion on this point, based on more concrete examples. 

Let us now consider the infinitesimal deformations $\delta\Re T$ of $\Re T$. They must satisfy (\ref{susy2bis}) and thus there must exist a compensating RR deformation $\delta C$ such that
\bea
\d\delta\Re T+\calj\cdot\d\delta C=0\ . 
\eea
This is possible if and only if
\bea\label{Tcond}
\d\d^\calj \delta\Re T\equiv 2i\del\delbar\delta\Re T= 0\ ,
\eea
which can be obtained directly from (\ref{tbianchi}). Indeed, by the $\d\d^\calj$-lemma (\ref{forlemma1}), (\ref{Tcond}) implies that we can write $\d^\calj\delta \Re T=\d\d^\calj\chi$, for some real $\chi\in U_1\oplus U_{-1}$, and such a deformation can be compensated by an RR-deformation $\delta C=\calj\cdot\d\chi$, up to an additional closed form that can be considered as part of the pure RR moduli. Notice that $(\calj\cdot\d\chi)_{0}=0$ and so the RR compensating shift belongs to $U_2\oplus U_{-2}$. More explicitly $\delta C=2i(\del\chi_{1}-\delbar\chi_{-1})$.  One can easily see that, by defining 
 \bea\label{Tprime}
 \hat\delta\Re T:=\delta \Re T-\calj\d^\calj\chi=\delta \Re T-2(\partial\chi_{1}+\delbar\chi_{-1})
 \eea we are led to 
 \bea\label{hatt}
 \d \hat\delta\Re T=0
 \eea
  and this condition is left unchanged under 
\bea\label{Tgauge}
\delta\Re T\rightarrow \delta\Re T+(\d\Lambda)_0
\eea 
for generic real $\Lambda\in U_{1}\oplus U_{-1}$. Indeed, (\ref{Tgauge}) corresponds to $\chi\rightarrow \chi-\Lambda/2$ (up to a $\d\d^\calj$-closed term) and thus 
\bea
\hat\delta\Re T\rightarrow \hat\delta\Re T+\d\Lambda\ .
\eea
Thus the deformations $\delta \Re T$ satisfying (\ref{Tcond}), modded out by the symmetry (\ref{Tgauge}) of the equation (\ref{susy2bis}), are identified by real elements of $\H^{0}_H(M;\mathbb{R})$ or, using (\ref{cohocond}), by $\H_H^{\rm ev}(M;\mathbb{R})$. 

A key point is that the symmetry (\ref{Tgauge}) is generically violated by the condition (\ref{dflat}).\footnote{Actually, this statement requires some reasonable non-degeneracy conditions, as can be seen from the argument given in appendix \ref{fluxhitchin}. However, as shown below, this assumption seems to be indirectly ensured by the requirement of having a consistent low-energy effective theory.} This is consistent with the interpretation  of (\ref{susy2bis}), and thus (\ref{hatt}),  as F-flatness condition and (\ref{dflat}) as D-flatness condition associated exactly to the RR gauge transformations (\ref{rrgauge}) \cite{lucapaul2}, since the symmetry (\ref{Tgauge}) can be seen as the imaginary extension of the RR gauge transformation ---  see below. Thus, following the usual approach in $N=1$ supersymmetric field theories, the D-flatness condition is taken into account by modding out the symmetry (\ref{Tgauge}) and thus the $\Re T$ deformations can be identified with $\H_H^{\rm ev}(M;\mathbb{R})$. Notice that all the equations we are considering  are preserved by the group of generalized diffeomorphisms $\calg_\calz\subset \calg$ [see (\ref{resym}) and (\ref{gendiff})], which generates deformations of $\Re T$ which are trivial in $\H_H^{0}(M;\mathbb{R})$. Thus, the D-flatness condition does not completely fix the representatives of the classes in $\H_H^{0}(M;\mathbb{R})$ describing the deformations of  $\Re T$ .

From the four-dimensional point of view, the deformations of $\Re T$ combine with the RR moduli, giving the lowest component of chiral fields.  Consistency with an effective $N=1$ low-energy description then implies that, since the RR moduli are unobstructed,  the NS infinitesimal deformations $\delta \Re T$ are unobstructed too. Thus, the finite deformations of $\Re T$ can be identified with an open subset of  
\bea\label{tm}
\calm_T\simeq \H_H^{\rm ev}(M;\mathbb{R})\ , 
\eea
at least in absence of D-branes that can gauge the RR axionic shift and generate D-terms for the $\Re T$ moduli ---  see section \ref{Dtermmoduli}. Combining (\ref{tm}) and (\ref{rrm}), we conclude that the $\calt$-moduli space can be locally identified as
\bea\label{Tmoduli}
\calm_\calt\simeq \H_H^{\rm ev}(M)\ ,
\eea
or better as a torus fibration 
\bea
0\rightarrow\calm_{\rm RR}\rightarrow \calm_\calt \rightarrow \calm_{T}\rightarrow 0\ .
\eea

To emphasize the complex structure of $\calm_\calt$, we can revisit its derivation given above in terms of real polyforms directly in terms of the complex polyform $\calt$. Let us consider the complex $\calt$-fluctuations defined in (\ref{gde}). From (\ref{susy2}) we see that they must satisfy the conditions
\bea\label{defFterm0}
\bar\partial \delta\calt_0+\del\delta\calt_{-2}=0\quad,\quad \bar\partial \delta\calt_{-2}=0\ .
\eea
Using (\ref{cohocond})  and the second condition in (\ref{defFterm0}), one can write $\partial\delta\calt_{-2}=\partial\bar\partial\chi_{-1}$. Then, defining 
\bea\label{tprime}
\hat\delta\calt_{0}:=\delta\calt_{0}-\partial\chi_{-1}\quad,\quad \hat\delta\calt_{-2}:= \delta\calt_{-2}
\eea we can write the above conditions as
\bea\label{defFterm}
 \bar\partial\hat\delta\calt_{0}=0\quad,\quad\bar\partial\hat\delta\calt_{-2}=0\ .
\eea

As above, to identify the physically inequivalent fluctuations, one has to mod out the RR gauge transformations and impose the D-flatness condition (\ref{dflat}). In this complexified language, an RR gauge transformation\footnote{Starting from a generic gauge transformation $C\rightarrow C+\Lambda$, with $\Lambda$ any polyform (of appropriate parity), then $\delta_\Lambda C_{-2}=\partial \Lambda_{-3}+\bar\partial \Lambda_{-1}$ and $\delta_\Lambda C_{0}=\partial \Lambda_{-1}+\bar\partial \Lambda_{1}$, where $\Lambda_{-1}=\Lambda_1^*$. But using the $\partial\bar\partial$-lemma we have $\partial \Lambda_{-3}=\bar\partial\partial \alpha_{-2}$ and thus we can rewrite the most generic gauge transformation as in (\ref{rrgauge}). Instead of the $\del\delbar$-lemma, we could also use only the condition $\H^{-2}_{\delbar}(M)=0$ to write $\partial \Lambda_{-3}=\bar\partial\alpha_{-1}$. Also in this case, the residual symmetry generated by $\Lambda_{-3}$ would not affect the conclusions obtained by considering just the symmetry (\ref{rrgauge}).}
 \bea\label{rrgauge}
 \delta_\Lambda C_{-2}=\bar\partial \Lambda_{-1}\quad,\quad\delta_\Lambda C_{0}=\partial \Lambda_{-1}+\bar\partial \Lambda_{1}\quad,\quad \text{with}\quad\Lambda_{-1}=\Lambda_1^*\ ,
 \eea 
acts on $\hat\delta\calt$ in the following way  
\bea\label{caltdef}
\hat\delta\calt_{0}\rightarrow \hat\delta\calt_{0}-i\bar\partial\Lambda_{1}\quad,\quad
\hat\delta\calt_{-2}\rightarrow \hat\delta\calt_{-2}-i\bar\partial\Lambda_{-1} \ .
\eea
Notice that, although the RR gauge transformation corresponds to a {\em real} $\Lambda:=\Lambda_{1}+\Lambda_{-1}$, the conditions (\ref{susy2}) are in fact invariant for arbitrary {\em complex} $\Lambda$. In particular, a purely imaginary $\Lambda\rightarrow i\Lambda$ (with $\Lambda$ real)  corresponds to the transformation (\ref{Tgauge}) and directly shows the above statement that (\ref{Tgauge}) can be considered as the imaginary extension of an RR gauge transformation. As above, the moduli space is given by the deformations preserving the F-flatness conditions (\ref{defFterm}), modded out by the gauge transformations (\ref{rrgauge}), with $\Lambda$ complex. Thus, using (\ref{cohocond}), in this complexified formulation we get $\calm_\calt\simeq \H^0_{\delbar}(M)$, and thus (\ref{Tmoduli}). Notice that  this second derivation suggests that explicit use of the $\del\delbar$-lemma could be avoided. Thus it is conceivable that, under suitable conditions, the $\del\delbar$-lemma could be relaxed without substantially changing the conclusions of our analysis. Nevertheless, for simplicity, we will continue assuming it in the following.

We see that the condition (\ref{cohocond}) allows a completely topological characterization of the $\calt$-moduli and thus of the corresponding chiral fields. More explicitly, we can fix a certain reference $\calt^{0}$ and write $\calt=\calt^{0}+\Delta \calt$ where  $\Delta\calt$ is a finite deformation associated with a certain twisted cohomology class $[\hat\Delta \calt]$ in $\H^{\rm ev}_H(M)$, which is the integrated finite version of (\ref{tprime}). The 4D chiral fields $t^a$ are identified by expanding 
\bea\label{caltexp}
[\hat\Delta\calt]=t^a[\omega_a]\ ,
\eea 
where $[\omega_a]$ is a certain moduli-independent basis for $\H^{\rm ev}_H(M;\mathbb{R})$. We can then split
\bea\label{tsplit}
t^a=s^a+i c^a\ ,
\eea
where  $c^a$ are the RR moduli  and $s^a$  can be identified with the NS moduli encoded in $\Re T$.


\section{The dual picture: linear multiplets}
\label{linearsection}

We have identified the chiral multiplets $t^a$ of the 4D effective description with the deformations of the polyform $\calt$, which contains the moduli of $\Re T$ and $C$. However, one can  look for a dual parametrization of the degrees of freedom contained in $\Re T$ in terms of the polyform $\Im T$, which indeed contains the same information \cite{hitchin}. From (\ref{dflat}) we know that $e^{2A}\Im T$ must be closed and so it is natural to guess that the space of allowed deformations is still given by $\H^{\rm ev}_H(M;\mathbb{R})$. (Actually, in the presence of orientifolds, $[e^{2A}\Im T]$ has orientifold parity opposite to $[\hat\Delta\calt]$, cf.~appendix \ref{app:orientifold}.) This can be made more precise by saying that the $U_0$-representative in  $\H^{\rm ev}_H(M;\mathbb{R})$ must be fixed by (\ref{susy2bis}) or (\ref{tbianchi}), up to the action of the symmetry group (\ref{resym}). This was already suggested in \cite{toma}, which presented an argument based on a Hitchin-like functional, and an analogous argument is described in appendix \ref{fluxhitchin}. Thus, let us expand 
\bea\label{explin}
[e^{2A}\Im T]=l_a[\tilde\omega^a]\ ,
\eea
where $[\tilde\omega^a]$ is a basis for $\H^{\rm ev}_H(M;\mathbb{R})$. In particular, we can choose a basis $[\tilde\omega^a]$ dual to the basis $[\omega_a]$ introduced in the previous section, i.e.\ such that
\bea\label{paireven}
\int_M\langle\omega_a,\tilde\omega^b\rangle=\delta_a{}^b\ .
\eea
The parameters $l_a$ can be seen as 4D scalar fields belonging to linear multiplets. The other bosonic fields in these linear multiplets are given by 4D two-forms $B_a$ obtained by expanding the RR-gauge potentials with two 4D indices in appropriately defined harmonic representatives of $\tilde\omega^a\in\H^{\rm even}_H(M;\mathbb{R})$.

Recall our assumptions that the $z^I$ moduli (up to the conformal compensator $Y$) are completely lifted by the fluxes and D-branes do not play any role. Then, splitting  the $t^a$ moduli as in (\ref{tsplit}), we generically have
\bea
l_a=l_a(Y,\bar Y,s^b)=|Y|^2\hat l_a(s^b)\ ,
\eea
where we have also explicitly indicated how $l_a$ depends on the conformal compensator $Y$, which is fixed by the fact that the chiral fields $t^a$ have Weyl weight zero. As we will see, knowing $\hat l_a(s^b)$ allows us to write a set of equations determining  the K\"ahler potential. Unfortunately, the computation of the explicit functional dependence of  $\hat l_a(s^b)$ may be cumbersome. 

Notice that the function $l_a(Y,\bar Y,s^b)$ (or equivalently $\hat l_a(s^b)$) is not necessarily invertible to $s^a=s^a(l_a/|Y|^2)$. Indeed, no-scale models \cite{noscale} are characterized by a non-invertible relation \cite{toinenoscale}, as we will discuss in more detail in section \ref{sec:kaehler}.


\section{Domain walls, strings, instantons and holomorphic couplings}
\label{probes}

The above characterization of the parameters $z^I,t^a,l_a$ extracted from $\calz$, $\calt$ and $e^{2A}\Im T$ in terms of $H$-twisted cohomology classes  agrees very well with the 4D interpretation of different D-brane configurations. In particular, the use of $H$-twisted cohomologies together with the condition (\ref{cohocond}) is crucial in order to get a completely topological characterization  of the dependence on the closed string moduli of  tensions, charges and couplings of the 4D effective objects.

Let us start with a BPS domain wall, obtained by wrapping a D-brane on an internal generalized cycle $(\Sigma,{\rm F})$. Then, in our units $2\pi\sqrt{\alpha^\prime}=1$, the tension of the domain wall is given by \cite{lucal,lucasup}
\bea
\tau_{{\rm DW}}=2\pi\left|\int_M\langle \calz, j_{\rm DW}\rangle\right|\ ,
\eea
where $j_{\rm DW}$ is the generalized current associated with $(\Sigma,{\rm F})$ as in (\ref{Dcurrent}). It defines an integral element of $\H^{\rm od}_H(M;\mathbb{R})$ and thus we can expand $[j_{\rm DW}]=n^I\alpha_I+m_{J}\beta^J$ and the domain wall tension is
\bea
\tau_{\rm DW}=2\pi|m_{I}z^I +n^I\calg_I(z) |\ .
\eea
We see that its dependence on $z^I$ can be completely identified in terms of real twisted cohomology classes, without the need of any additional structure.   

An analogous discussion can be repeated for D-strings $\gamma\subset X_4$ obtained by wrapping a D-brane on an internal BPS generalized cycle, for which we can write $[j_{\rm string}]=n^a\omega_a$. The associated 4D effective action is \cite{lucal,lucasup}
\bea
S_{\rm string}=-2\pi\, n^a\int_{\gamma} \d^2\sigma\, l_a\sqrt{-\det g_{(4)}|_{\gamma}}+2\pi\, n^a \int_{\gamma} B_{a} \ ,
\eea 
where $B_a$ is the 4D two-form belonging to the same linear multiplet as $l_a$.
Notice that the BPS condition imposes $j_{\rm string}\in U_0$. Thus, had we not assumed (\ref{cohocond}), we would have lost the possibility to characterize the D-string action purely in terms of the topology of $(\Sigma,{\rm F})$ wrapped by the D-brane. 
   
Consider now a D-brane instanton. Also in this case, the BPS condition implies that $j_{\rm inst}\in U_0$ \cite{lucapaul2} and the on-shell action is given by
\bea
S_{\rm inst}=2\pi\int_M\langle \calt, j_{\rm inst}\rangle\ . 
\eea
Since $\d j_{\rm inst}=0$, writing $\calt=\calt^{0}+\Delta \calt$ for some fixed $\calt^{0}$, we can actually substitute $\Delta\calt$ with $\hat\Delta\calt$ (defined in (\ref{tprime})) in  $S_{\rm inst}$.  Clearly  $S_{\rm inst}$ depends only on the class of $\hat\Delta \calt$ in $\H^0_{\delbar} (M)\simeq \H^{\rm ev}_H (M)$. Expanding $[j_{\rm inst}]=n_{a}[\tilde\omega^a]$ and $[\hat\Delta \calt]$ as in (\ref{caltexp}), the on-shell instanton action can be written as
\bea\label{instact}
S_{\rm inst}=S^{0}_{\rm inst}+2\pi\, n_at^a\ ,
\eea
where $S^{0}_{\rm inst}$ does not depend on $t^a$. As expected, the corresponding contribution
\bea\label{instcorr}
\sim e^{-2\pi\, n_{a}t^a}
\eea
to the path integral breaks the axial symmetry $t^a\rightarrow t^a+i\alpha^a$. 

Finally, in the internal space   a space-filling BPS D-brane is identical to a BPS instanton \cite{lucal,lucapaul2}.   The associated (classical) 4D holomorphic coupling $f(t)$ is identical to the instanton action (\ref{instact}), i.e.\ 
\bea\label{holcoupling}
f(t)=2\pi\int_M\langle \calt, j_{\text{space-filling}}\rangle= f^{0}+2\pi\, n_at^a\ .
\eea
As for D-brane instantons, the condition (\ref{cohocond})  and the use of $H$-twisted cohomologies is crucial to have a fully topological characterization of the dependence of $f(t)$ on the closed string moduli.


\section{The K\"ahler potential}
\label{sec:kaehler}

In the previous sections we have characterized  the low-energy spectrum of massless  chiral  fields $t^a$, and their dual linear multiplets $l_a$, in purely geometrical/topological terms. In particular, the non-exhaustive control over the analytical properties of the internal geometry has been supplied by four-dimensional consistency arguments. I would now like  to complete the above results by discussing the effective K\"ahler potential of the low-energy effective action. This will also give further support  to the above picture. 

As stressed in \cite{lucapaul2}, warped flux compactifications are very naturally described in terms of 4D superconformal theories. The reason is that in this formulation one can use directly the 4D-metric $\d s^2_{X_4}$ appearing in (\ref{metricsplit}) as the dynamical one, without having to rescale it from the beginning to go to the Einstein frame. This permits a more direct comparison between the four-dimensional and ten-dimensional  pictures.  

Let us continue working with the simplifying assumption that we have only moduli/chiral fields $t^a$ and no  D-branes.  Then, at the classical level, the scalar sector of the effective theory must be completely specified in terms of a conformal K\"ahler potential $\caln(Y,\bar Y,t,\bar t)$, where $Y$ is the conformal compensator.  Let us recall what are its basic features, derived from purely 4D arguments (see e.g.\ \cite{toine} for more details). The superconformal Lagrangian is given by 
\bea
\call=-3\int\d^4\theta\, \caln\ ,
\eea
where $\d^4\theta$ is a formal way of writing the full superspace measure in supergravity. This produces an Einstein term of the form 
\bea\label{einsteinaction}
\call=\frac12\,\caln\, \calr+\ldots\quad.
\eea
Since $\caln$ must have Weyl weight two and the chiral fields $t^a$ have Weyl weight zero,  the dependence of $\caln$ on $Y$ is fixed to be of the form  
\bea\label{calnsplit}
\caln=|Y|^2\hat\caln(t,\bar t)\ .
\eea
Then, the usual Einstein-frame K\"ahler potential is given by
\bea\label{einkaehler}
\calk=-3\log \hat\caln(t,\bar t)\ .
\eea 

The Einstein-frame action is obtained by gauge-fixing the superconformal action, in particular by imposing the condition $Y=M_{\rm P}e^{\calk/6}$ (where $M_{\rm P}$ is the 4D Planck mass), which breaks the complexified Weyl invariance.
From (\ref{einsteinaction}) it is clear that this condition leads to the Einstein frame since it corresponds to imposing $\caln=M^2_{\rm P}$.

For our purposes it is important to recall how linear multiplets are obtained by a duality transformation in the superconformal framework \cite{linear}. First, one has to assume that the K\"ahler potential has  the form
\bea
\caln=\caln(Y,\bar Y,s)
\eea 
where $s^a=(t^a+\bar t^a)/2$. Then,  the dual linear multiplets $l_a$ are given by a Legendre transformation
\bea\label{linearchiral}
l_a=\frac{3}{4\pi}\,\frac{\del\caln}{\partial s^a}\ ,
\eea
which can be formally considered not only as a full superfield equation but also as its lowest component involving bosonic scalar fields, as we will do in the following.

Let us now go back to our flux compactifications and their moduli, as described in the previous sections. First, the conformal K\"ahler potential is univocally determined by dimensionally reducing the 10D supergravity action and comparing it with (\ref{einsteinaction}). The resulting $\caln$ depends only on the NS fields and thus can be expressed  in terms of the pure spinors $\calz$ and $T$   as follows \cite{lucapaul2}:
\bea\label{geomkaehler}
\caln=\frac{i\pi}2\int_M\langle \calz,\bar\calz\rangle^{1/3}\langle T,\bar T\rangle^{2/3}\ .
\eea
This expression is completely fixed by supersymmetry and, in fact, can be considered a sort of microscopical K\"ahler potential giving, together with the superpotential (\ref{totsup}), the full set of 10D supersymmetry equations (including those for AdS$_4$-compactifications) \cite{lucapaul2}. Once we have $\caln$, in order to obtain the usual Einstein-frame K\"ahler potential as described above,  one has  first to isolate a conformal compensator $Y$ from $\calz$ by choosing a reference $\calz^{0}$ as in (\ref{fixedz}). Then
\bea\label{einkaehler2}
\calk=-3\log \hat\caln=-3\log\big(\frac{i\pi}{2}\int_M\langle \calz^{0},\bar\calz^{0}\rangle^{1/3}\langle T,\bar T\rangle^{2/3}\big)\ .
\eea
The problem is now that, in order to consider (\ref{geomkaehler}) and (\ref{einkaehler2}) as effective low-energy K\"ahler potentials for the massless moduli, one needs to extract the explicit dependence of $\caln$ on $t^a$ and $\bar t^{a}$. This is not trivial since $\caln$ does not have a simple interpretation in terms of the topological data characterizing the moduli, differently from what happens for the usual K\"ahler potentials in ordinary Calabi-Yau compactifications.

However, the situation is better if one considers the derivatives of $\caln$. First, notice that the conformal K\"ahler potential (\ref{geomkaehler}) depends only on NS fields and thus can depend on $t^a$ only through its real part $s^a=(t^a+\bar t^a)/2$.
The  first order variation of $\caln$ under a general variation of $\delta\Re T\in U_0$ is given by
\bea\label{1var}
\delta\caln=\frac{4\pi}{3}\int_M\langle \delta\Re T,e^{2A}\Im T\rangle\ .
\eea
Now, we can restrict to moduli deformations discussed in section \ref{caltmoduli}. They are given by deformations $\delta\Re T$ satisfying (\ref{Tcond}), which allow to define the closed form $\hat\delta \Re T$ as in (\ref{Tprime}). Shifting $\hat\delta \Re T$ by an exact polyform corresponds to the transformation (\ref{Tgauge}), which is eventually fixed by the D-flatness condition (\ref{dflat}). The key-point is that, because of (\ref{dflat}) and the fact that $e^{2A}\Im T\in U_0$, we need only care about the class of  $\hat\delta\Re T$ in $\H^{\rm ev}_H(M;\mathbb{R})$. More explicitly, if we write
\bea
[\hat\delta\Re T]=\delta s^a[\omega_a]
\eea
then
\bea\label{derN}
\frac{\partial\caln}{\partial s^a}=\frac{4\pi}{3}\int_M\langle \omega_a,e^{2A}\Im T\rangle=\frac{4\pi}{3}\,l_a(Y,\bar Y,s)\ ,
\eea
where in the last equality we have used the expansion (\ref{explin}) with the choice (\ref{paireven}). Thus, we see that the K\"ahler potential (\ref{geomkaehler}) and the characterization of the moduli in terms of chiral and linear multiplets given in sections \ref{caltmoduli} and \ref{linearsection}, are in perfect agreement  with (\ref{linearchiral}), which is expected from purely 4D arguments. 

These results suggest an alternative way to extract  from (\ref{geomkaehler}) the explicit dependence of the effective conformal K\"ahler potential on $s^a=(t^a+\bar t^a)/2$. One should first find the explicit functional dependence of $l_a(Y,\bar Y,s)$, which encodes the dependence of the twisted cohomology class defined by $e^{2A}\Im T$ in terms of the twisted cohomology class defined by $\Re T$. This must be computed from the specific six-dimensional internal geometry as described in sections \ref{caltmoduli} and \ref{linearsection}. Then, since the dependence of $\caln$ on $Y$ has  the form (\ref{calnsplit}),   by integrating (\ref{derN}) one can obtain $\caln$ up to an additional integration constant which can be fixed by evaluating (\ref{geomkaehler}) on the specific configuration corresponding to the initial conditions. Isolating the compensator as in (\ref{calnsplit}), on can obtain the Einstein-frame K\"ahler potential (\ref{einkaehler}).

Finally, notice that although we have for simplicity assumed that all the $\calz$-moduli $z^I$, apart for the compensator $Y$, are lifted by the superpotential (\ref{supeff}), the expressions (\ref{geomkaehler}) and (\ref{einkaehler2}) are generically valid even in presence of residual  moduli $z^i$ (which would parametrize $\calz^0$). In particular,  (\ref{derN}) is still valid but now on the right-hand side one has $l_a=l_a(z,\bar z,s)$ and thus the linear multiplets also depend on the $\calz$-moduli. Again, computing  $l_a=l_a(z,\bar z,s)$ from the internal geometry provides a way to clarify the structure of $\caln$, and thus of $\calk$, by integrating (\ref{derN}). However, in this case, the integration `constant' would generically depend on the $\calz$-moduli $z^I$ and its explicit form needs to be determined by other means.  

\bigskip

At this point, it may be clarifying to discuss the limit in which one assumes a constant warp-factor. First of all, considering $e^{2A}$ constant, from (\ref{dilatonwarp}) one can easily see that the conformal K\"ahler potential (\ref{geomkaehler}) can be factorized as follows
\bea\label{facn}
\caln_{\rm unwarped}=\frac{\pi i}{2}\left(i\int_M\langle\calz,\bar\calz\rangle\right)^{1/3}\left(i\int_M\langle T,\bar T\rangle\right)^{2/3}\ .
\eea
If we set $\calz=Y^3\calz^0$, we are led to the K\"ahler potential (up to an additional constant)
\bea\label{unkaehler}
\calk_{\rm unwarped}=-\log\left( i\int_M\langle\calz^0,\bar\calz^0\rangle\right)-2\log \left(i\int_M\langle T,\bar T\rangle\right)\ .
\eea
In the SU(3)-structure case, $\calk_{\rm unwarped}$ is the generalized K\"ahler potential obtained  in \cite{grimm}, which can be seen as an orientifold truncation of the $N=2$ K\"ahler potentials of \cite{granalouis}, where $\calz_0$ and $T$ belong to vector- and hyper-multiplets, respectively. However, notice that the factorization of $\caln_{\rm unwarped}$ into (\ref{facn}), which leads to the split of $\calk_{\rm unwarped}$ typical of an underlying $N=2$ structure,  seems  possible only if the warping is constant. 

Indeed, in the unwarped approximation, the results of \cite{lucapaul2} we started from reproduce those of  \cite{granalouis,grimm}, which in turn provide a unifying formulation of different results present in the literature on unwarped compactifications. A detailed discussion can be found e.g.\  in  \cite{granareview}. So, let us just briefly comment on it. We are interested in low-energy effective potentials, where very massive modes are integrated out. Using the untwisted picture [cf.~appendix \ref{twist}] for clarity, the external Einstein and dilaton equations imply that
\bea\label{warpeq}
\nabla^m(e^{-2\Phi}\nabla_m e^{4A})=e^{4A}F^2+\rho^{\rm loc}\ ,
\eea
where $\rho^{\rm loc}$ is the energy density associated with the localized sources. Thus, in order to consider the warping as approximately constant one needs the right-hand side of (\ref{warpeq}) to be very-small or, in other words, the RR-fluxes may be considered as a small perturbation, let us say of order $\varepsilon$, of some underlying supersymmetric vacuum. Let us assume that it is the case and try to expand all the equations in $\varepsilon$.

First, supersymmetry imposes that \cite{pauldimi}
\bea
e^{4A-2\Phi}*H=-e^{3A}[\sigma(F)\wedge \Re T]_{\it 3}+\d(...)\ ,
\eea  
where the operator $\sigma$ is defined below (\ref{mukai}). Then on a compact manifold, since we are assuming $\d H=0$,  $H$ must be vanishing at zeroth order in $\varepsilon$.  This implies that at zeroth order the pure spinors satisfy the equation
\bea\label{lweq}
\d\calz=0\quad,\quad \d T=0\ .
\eea
and thus describe a vacuum with (at least) $N=2$ supersymmetry and no $H$-field. \footnote{Using the terminology of \cite{gualtieri}, $\calz$ and $T$ satisfying (\ref{lweq}) define a generalized CY metric.}  In the SU(3)-structure case, we have
\bea\label{purespinorCY}
\text{IIA}&:& T=e^{-\Phi}\,\Omega \quad,\quad \calz=e^{3A-\Phi}e^{iJ+B}\ ,\cr
\text{IIB}&:& T=e^{-\Phi}\,e^{iJ+B}\quad,\quad \calz=e^{3A-\Phi}\Omega\ ,
\eea  
and then (\ref{lweq}) imply  that $\Omega$ and $J$ describe an ordinary CY manifold with constant dilaton. In this case the condition (\ref{cohocond}) is indeed automatically satisfied and the spectrum discussed in this paper (at zeroth order, i.e.\ for $H=0$) boils down to the orientifolded scalar spectrum of \cite{louis}. In this approximation, the $H$-twisting is a perturbative effect and shows up in an effective potential.\footnote{For example,  the (untwisted) $\calt$-moduli generically get D-terms and F-terms. The first  are produced by the gauging of the (untwisted) RR shift symmetry that involves $[H]$-exact forms.  The latter can be obtained directly from (\ref{totsup}), rewritten in untwisted picture by replacing $\d$ with $\d_H$ and truncated according to the zeroth-order truncation.}  The same is true for other perturbative effects, like the so-called `geometric' fluxes that describe the deviation, measured by torsion classes, of the metric from the Ricci-flat one. The  K\"ahler potential  (\ref{unkaehler}) gives exactly the K\"ahler potential obtained in \cite{louis}. See \cite{granareview} for more details.

\section{No-scale models}
\label{sec:noscale}

As already stressed, in the superconformal approach, the dependence  of the linear multiplets $l_a(Y,\bar Y,s^a)$ (with $s^a=\Re t^a$)  on the dual chiral multiplets $t^a$ is not generically required to be invertible in terms of the $s^a$ and in fact the interesting case of no-scale supergravities \cite{noscale} is obtained  when it is not or, in other words, when the matrix
\bea
h_{ab}=-\frac{\partial^2\caln}{\partial s^a\partial s^b}
\eea
is degenerate \cite{toinenoscale,barbieri}. Indeed, using (\ref{calnsplit}) and (\ref{einkaehler}) one can easily compute 
\bea
h_{ab}=\frac13\caln(\calk_{ab}-\frac13 \calk_a\calk_b)\ ,
\eea
where we are considering $\calk$ as a function of the real coordinates $s^a$,
$\partial_a\calk:=\partial\calk/\partial s^a$ and $\calk_{a b}:=\partial^2\calk/(\partial s^a\partial s^{b})$. Then, denoting with $\calk^{ab}$ the inverse of $\calk_{a b}$, we have
\bea
\det h_{ab}=\det(\frac13\caln\calk_{ab}) ( 1-\frac13 \calk_c\calk^{cd}\calk_d)\ , 
\eea
and imposing degeneracy of $h_{ab}$ leads to the well-known no-scale condition 
\bea\label{noscalecond}
\calk_a\,\calk^{ab}\,\calk_b=3\ . 
\eea

Let us for example consider the case in which we can identify one of the chiral fields $t^a$, let us call it $\rho=r+ic_r$ and denote with $\phi^\alpha$ the remaining chiral fields, such that $\partial_r^2\caln=0$ and $\partial_r\del_{\alpha}\caln=0$. Then, we can write
\bea
\calk=-3\log [\rho+\bar\rho+f(\phi+\bar \phi)]+\text{const.}
\eea

At the level of twisted cohomology classes, the no-scale property can be interpreted as the existence of a trajectory in $\H_H^{\rm ev}(M;\mathbb{R})$, parametrizing the deformations of $\Re T$, which leaves the class of $e^{2A}\Im T$ in $\H_H^{\rm ev}(M;\mathbb{R})$ unchanged. As we will see in section \ref{example:wCY}, an explicit example of this mechanism is provided by the well known type IIB warped CY compactifications \cite{gp,gubser,gkp}.  It would be interesting to clarify the relation between the 10D interpretation given above of the 4D no-scale condition  and the 10D conditions for the class of generalized non-supersymmetric  vacua found in \cite{nonsusy}.

Finally, notice that by defining $s^A:=(s^0,s^a)$, with $s^0:=\log |Y|^2$, and $l_A:=(3/4\pi)\partial_A\caln$, then $l_A(s^B)$ is required to be invertible, as follows from the 4D  requirement that the metric obtained from $\caln$ must be non-degenerate \cite{toine}. Performing a duality transformation involving also the conformal compensator would lead to a new-minimal supergravity  \cite{newminimal} involving only linear multiplets, as for example discussed  in \cite{ferraradauria} in the context of the effective five-brane theory on CY spaces.  



\section{D-term moduli-lifting from D-brane gauging}
\label{Dtermmoduli}

In most of the above discussions we have for simplicity assumed no space-filling D-branes, arguing that the $\calt$-moduli remain classical unobstructed. This section is intended to add some comments on the effect of D-branes on the $\calt$-moduli, leaving a more detailed discussion for the future. 

The presence of space-filling D-branes changes the story  and not only because they enter the superpotential (\ref{supeff}) as described in section \ref{calz}. Indeed,  it is well known that D-branes can introduce D-terms for part of the closed string moduli, giving them a mass by the St\"uckelberg mechanism. Let us see how this works in our general setting.

Supersymmetric D-branes must obey the condition (\ref{braneDterm}). Let us introduce a new current $\hat \jmath_{\text{D-brane}}$ which has orientifold parity opposite  to $j_{\text{D-brane}}$ and expand it in cohomology as $[\hat \jmath_{\text{D-brane}}]=n^a[\omega_a]$. Then integration of (\ref{braneDterm}) over the internal manifold produces the condition\footnote{\label{invsign}The choice of $\hat \jmath_{\text{D-brane}}$ hides a subtlety. Indeed, using $j_{\text{D-brane}}$ in (\ref{braneDterm})  would give, after integration, an empty equation. The point is that the D-flatness condition (\ref{braneDterm}) is local and must be satisfied by both the D-brane and its orientifold image. Thus, in the covering space, one has to  replace the D-brane image with the anti-brane image to get the correct result. This is confirmed by the derivation of the D-term from the gauging of the RR-axion given later in this section.} 
\bea\label{4Ddflat}
n^al_a(Y,\bar Y,t+\bar t)=0\ .    
\eea
This condition implies that the presence of D-branes naturally leads to a lifting of the $t^a$ moduli. Notice that in no-scale models the dependence of $l_a$ on $t^a$ is not invertible and thus there will be at least one remaining modulus $t^a$ unlifted. 

It is easy to see that (\ref{4Ddflat}) can be interpreted as the D-flatness condition coming from the gauging of an RR axionic symmetry under the D-brane U(1) gauge group. Indeed, from the Bianchi identity (\ref{BI}) one can see that a D-brane U(1) gauge transformation parametrized by $\lambda$ induces a  shift of the RR-potential $C\rightarrow C-\lambda \hat \jmath_{\text{D-brane}}$.\footnote{This can be seen as follows. Using the current $\theta$ introduced in (\ref{fsplit}), we have $\d(F-\theta)=0$ and thus the RR gauge potential $C$ is defined by $\d C=F-\theta$. On the other hand, under a world-volume U(1) gauge transformation $A\rightarrow A+\d\lambda$, we have $\delta_\lambda\theta=\d(\lambda \hat \jmath_{\text{D-brane}})$ and thus $\delta_\lambda C=-\lambda \hat \jmath_{\text{D-brane}}$} Its imaginary extension is given by $\Re T\rightarrow \Re T+\lambda \hat \jmath_{\text{D-brane}}$ and produces a D-term which, using standard 4D supergravity formulas (see e.g.\ \cite{toine}), takes the form  
\bea\label{dlambda}
D_\lambda=-\frac32\,\delta_\lambda \caln=2\pi\int_M\lambda\langle e^{2A}\Im T, \hat \jmath_{\text{D-brane}}\rangle\ .
\eea
This exactly reproduces the D-term found in \cite{lucasup} starting from the D-brane effective action.  Notice that all the factors are completely fixed and then this nontrivial matching is possible only thanks to the peculiar $2/3$-power of $\langle T,\bar T\rangle $ appearing in (\ref{geomkaehler}). Clearly, since $\lambda$ is a generic function on the internal cycle wrapped by the D-brane, by imposing $D_\lambda=0$ one gets the condition (\ref{braneDterm}).

One can write $D_\lambda=\int_\Sigma\lambda\cald$,    using the  D-term density 
\bea\label{Dtermdensity}
\cald= 2\pi\big[(e^{2A}\Im T)|_\Sigma\wedge e^{\rm F}\big]_{\rm top}\ ,
\eea 
where $(\Sigma,{\rm F})$ is the internal generalized cycle wrapped by the D-brane.    One the ways  \cite{lucasup} to see that $\cald$ can be identified with a D-term density is by expanding the D-brane action around a supersymmetric vacuum, obtaining  an untruncated D-like term
\bea\label{dpot}
V_{\cal D}=\frac1{4\pi} \int_\Sigma \frac{\cald^2}{\big[ \Re T|_\Sigma\wedge e^{\rm F}\big]_{\rm top}}\ .
\eea

Notice now that, for $\lambda$ constant, the transformation $C\rightarrow C-\lambda \hat \jmath_{\text{D-brane}}$ is not  associated with a D-brane gauge transformation but can be seen as a gauging of the RR-axionic shift
\bea
t^a\rightarrow t^a+i\lambda n^a\ .
\eea 
The associated D-term is just 
\bea
D=D(Y,\bar Y,t,\bar t)=2\pi\int_M\langle e^{2A}\Im T, \hat \jmath_{\text{D-brane}}\rangle=-\frac32n^a\partial_a\caln\ ,
\eea
which, imposing $D=0$, gives  exactly the D-flatness condition (\ref{4Ddflat}). Moreover, by substituting in (\ref{dpot})  the zero-mode ansatz 
\bea\label{dtermtrunc}
\cald\ \rightarrow\ \frac{D\,\d \text{vol}_\Sigma}{\int_\Sigma \d \text{vol}_\Sigma}\quad,\quad  \big[ \Re T|_\Sigma\wedge e^{\calf}\big]_{\rm top}\ \rightarrow\ \frac{(\Re f)\,\d \text{vol}_\Sigma}{2\pi\int_\Sigma \d \text{vol}_\Sigma}
\eea 
where $\d \text{vol}_\Sigma$ can be any volume form on the internal cycle and $f$ is the D-brane holomorphic coupling given in (\ref{holcoupling}), one gets the expected effective 4D formula for the D-brane induced D-term potential for the $t^a$ moduli:
\bea
\label{u(1)dterm}
V_D=\frac12\, (\Re f)^{-1}D^2\ .
\eea


\section{A subcase: IIB warped CY compactifications}
\label{example:wCY}

As an example, let us consider the subcase of IIB warped CY  compactifications \cite{gp,gubser,gkp}. In addition to the obvious physical interest in this class of vacua, the motivation for this choice is simple: in this case there is an underlying CY structure which allows a considerable  simplification of the analysis, in particular disentangling the reciprocal dependence of $\Re T$ and $e^{2A}\Im T$ when they are imposed to solve the supersymmetry equations.  On the other hand, this class of vacua has the non-trivial feature of having a non-vanishing $H$ and thus allows an explicit application of the $H$-twisted cohomology classes that are at the base of the formalism presented in this paper. 

 In this section, it will be convenient to pass to the untwisted picture where one considers polyforms that are gauge-invariant under $B$-field gauge transformations and uses the twisted differential $\d_H:=\d+H\wedge$.  In this case, the associated twisted cohomology can be computed by considering first ordinary de Rham cohomology and then, in a second step, the $[H]$-cohomology associated with the operator $[H]\wedge$ acting on the de Rham cohomology classes.\footnote{Actually, in general, this procedure could hide subtleties related to the `formality' of the manifold. See \cite{cavalcanti} for a detailed discussion.} Notice that, as will be clearer from the following discussion, this approach is quite different in nature from the one usually adopted in the literature on warped CY compactifications, where the $H$ field is treated on the same footing as the RR fuxes and the relevant cohomology classes are the ones of the underlying CY space.

In these vacua the metric has the form
\bea
\d s^2=e^{2A}\d x^\mu\d x_\mu +g_s\,e^{-2A} \d s^2_{\text{CY}_3}\ ,
\eea
where $g_s=e^\Phi$, and the axion-dilaton $\tau=C_0+i/g_s$  is constant. The $H$ flux must be primitive with $H^{0,3}=0$ and the RR fluxes are given by the conditions $g_s F_{\it 3}=-*H$ and $g_sF_{\it 5}=-4*\d A$.
The underlying CY space is completely specified by the holomorphic $(3,0)$-form $\Omega_{\rm CY}$ and the K\"ahler form $J_{\rm CY}$. In particular, we choose the normalization of $\Omega_{\rm CY}$ to be fixed by the condition
\bea\label{normCYcond}
\Omega_{\rm CY}\wedge \bar\Omega_{\rm CY}=\frac{4i}{3}\,g_s\,J_{\rm CY}\wedge J_{\rm CY}\wedge J_{\rm CY} \ .
\eea
Finally, a compact background must include O3-planes.

Let us start with the $\calz$ pure spinor, which is simply given by 
\bea\label{calzwCY}
\calz&=&\Omega_{\rm CY}\ .
\eea
The generalized Hodge decomposition of the polyforms reads
\bea\label{ghodgecomplex}
U_3=\Lambda^{3,0}&&\quad,\quad U_2=\Lambda^{2,0}\oplus \Lambda^{3,1}\quad,\quad U_1= \Lambda^{2,1}\oplus\Lambda^{1,0} \oplus\Lambda^{3,2}\ ,\cr
&&U_0= \Lambda^{0,0}\oplus\Lambda^{1,1}\oplus \Lambda^{2,2}\oplus\Lambda^{3,3}\ ,
\eea  
and the others $U_{-k}$ (with $k > 0$) are obtained by complex conjugation $U_{-k}=\overline{U_k}$. In this case, the generalized Dolbeault  operator is given by
\bea
\delbar_{H}=\bar\partial+H^{1,2}\wedge\ ,
\eea
where $\delbar$ is the ordinary Dolbeault  operator associated with the CY complex structure. The $\delbar_H$-cohomology is isomorphic to the $[H^{1,2}]$-cohomology applied to the standard Dolbeault cohomology: 
\bea
\H^k_{H}(M) \simeq \bigoplus_{r-t=k}\H^{r,t}_H\quad,\text{ with}\quad
\H^{r,t}_H:=\frac{{\rm ker}[H^{1,2}]:\H^{r,t}\rightarrow \H^{r+1,t+2}}{\Im [H^{1,2}]:\H^{r-1,t-2}\rightarrow \H^{r,t}}\ .
\eea
Notice that $h^{0,2}=h^{1,3}=0$ because of the underlying CY geometry, and thus the condition (\ref{cohocond}) is automatically satisfied.\footnote{\label{tori}There could be vacua of this kind where  the underlying CY space has actually reduced holonomy but the fluxes are still sufficient to break the supersymmetry to $N=1$. In this case $h^{0,2}$ and $h^{1,3}$ could be non-vanishing and the condition (\ref{cohocond}) should be an effect of the $H$-twisting. A simple class of  vacua of this kind is described in \cite{kachrutorus} (see also \cite{polfrey}), where explicit examples on the  $T^6/\mathbb{Z}_2$ orientifold with O3-planes are given, in which $h^{0,2}_-=0$ but $h^{1,3}_+=3$. Although these models would require a separate discussion, because of their rich cohomological structure, it is interesting to see how $(\H^{1,3}_{+})_H$ vanishes for $N=1$ solutions of this kind, while it does not vanish for examples with supersymmetry enhanced to $N=2$. Strictly $N=1$ solutions given in \cite{kachrutorus}, based on factorized $T^6=T^2\times T^2\times T^2$, have $H^{1,2}\sim \epsilon_{ijk}\d z^i\wedge \d \bar z^j\wedge \d \bar z^k$, where $z^i$ are the complex coordinates on the three two-tori. Then, in these cases we clearly have $\H^{1,3}_+=[H^{1,2}]\wedge \H^{0,1}_-$ and thus $(\H^{1,3}_{+})_H=0$. On the other hand, the explicit $N=2$ example provided  in \cite{kachrutorus} has 
$H^{1,2}\sim (\d z^1\wedge \d \bar z^{\bar 2}\wedge \d \bar z^{\bar 3}+\d z^2\wedge \d \bar z^{\bar 3}\wedge \d \bar z^{\bar 1})$ and in this case we see that  $\dim(\H^{1,3}_{+})_H=1$ since the element $\d z^3\wedge \d \bar z^{\bar 1}\wedge \d \bar z^{\bar 2}\wedge \d \bar z^{\bar 3}\in\H^{1,3}_+$ is not $[H^{1,2}]$-exact.}

Le us first consider the $\calz$-moduli space. It is easy to see that 
\bea
\calm^{\rm wCY}_{\calz}\simeq \H^{\rm od}_H(M;\mathbb{R})_+\simeq\H^{3,0}_-\oplus (\H^{2,1}_-)_{H}\ ,
\eea
where $\H^{3,0}$ parametrizes the overall constant rescaling of $\Omega_{\rm CY}$, while $(\H_-^{2,1})_{H}$ parametrizes the complex structure deformations that do not violate the condition $H^{0,3}=0$.\footnote{In the $T^6/\mathbb{Z}_2$ orientifold models discussed in footnote \ref{tori} we still have $\calm_{\calz}\simeq\H^{3,0}_-\oplus (\H^{2,1}_-)_{H}$. Indeed, the O3 projection imposes $\iota^*\calz=\sigma(\calz)$ (see appendix \ref{app:orientifold}) and thus possible deformations of $\calz$ along $\Lambda^{1,0}\oplus\Lambda^{3,2}$, which would correspond to the so-called $\beta$- and $B$-deformations, are associated to the $[H]$-twisting of $\H^{1,0}_+$ and $\H^{3,2}_+$, which both vanish on $T^6/\mathbb{Z}_2$.}

The effective superpotential  (\ref{supeff}) becomes
\bea\label{wCYsupeff}
\calw_{\rm eff}=\int_M\langle \Omega_{\rm CY}, F_{\it 3} \rangle\ ,
\eea
where  $\calw_{\rm eff}$ should be thought of as a holomorphic function on $\calm^{\rm wCY}_{\calz}$. Since the Mukai pairing is non-degenerate, it is immediate to see that imposing $\d\calw_{\rm eff}=0$ in $\calm^{\rm wCY}_{\calz}$ is equivalent to imposing that the complex structure is such that $[F^{0,3}]=0$ and $F^{1,2}$ is trivial in $\delbar_{H}$-cohomology. This means that (in ordinary cohomology)  $[F^{1,2}]=c[H^{1,2}]$ for some constant $c$, which will eventually be identified with $i/g_s$.

The other pure spinor $T$ takes the form
\bea\label{wT}
T &=& g^{-1}_s\exp(ig_se^{-2A}J_{\rm CY}+B)\ ,
\eea
where $B$ is a $(1,1)$-form.\footnote{To be precise, in order to keep track of the $B$-field degrees of freedom, here we are actually using a `mixed-twisted' picture where one splits $H=H_0+\d B$ and uses the twisted differential $\d_{H_0}$, whose cohomology is however isomorphic to the one computed from $\d_H$.}  


Recalling that in this case the O3 projection acts as $\iota^*\calt=\sigma(\calt)$ and $\iota^*\Im T=-\sigma(\Im T)$ (see appendix \ref{app:orientifold}), the corresponding cohomology classes $\H^{\rm ev}_+(M)$ and $\H^{\rm ev}_-(M)$ are given by 
\bea\label{wCYcc} 
[\hat\Delta\calt]\in\H^{\rm ev}_H(M)_+\simeq(\H^{0,0}_+\oplus\H^{1,1}_-\oplus \H^{2,2}_+\oplus \H^{3,3}_- )_{H}\simeq\H^{1,1}_-\oplus \H^{2,2}_+\ ,\cr
 [e^{2A}\Im T]\in \H^{\rm ev}_H(M)_-\simeq(\H^{0,0}_-\oplus\H^{1,1}_+\oplus \H^{2,2}_-\oplus \H^{3,3}_+ )_{H}\simeq\H^{1,1}_+\oplus \H^{2,2}_-\ .
\eea
Notice that $\hat\Delta \calt_{\it 6}$ already vanishes in (orientifolded) cohomology, while $\hat\Delta \calt_{\it 0}$ does not vanish in ordinary cohomology but is `non-closed' in $[H]$-cohomology. Thus in our approach $\hat\Delta \calt_{\it 0}$, which would be the axion-dilaton modulus in an ordinary Calabi-Yau compactification, is removed from the spectrum of the $H$-twisted cohomology. This is why the axion-dilaton is not present in the effective superpotential (\ref{wCYsupeff}), differently from what happens in the Gukov-Vafa-Witten superpotential (\ref{gvw}). On the other hand, $\H^{3,3}_+$ is not present in $\H^{\rm ev}_H(M)_-$, because it is trivial in the $[H]$-cohomology. The simplest way to see this is by noticing that $(\H^{3,3}_+)_{H}$ is Poincar\'e dual to $(\H^{0,0}_+)_{H}$, which vanishes.

We can expand the $\calt$-moduli as follows 
\bea\label{defwcy}
[\hat\Delta\calt]=\phi^a[\chi_a]+t^A[\omega_A]\in \H^{1,1}_-\oplus \H^{2,2}_+\ ,
\eea
where $\chi_a$ and $\omega_A$  are bases  for $\H^{1,1}_-$ and $\H^{2,2}_+$, respectively. On the other hand, we can expand $e^{2A}\Im T$ in the dual cohomology basis  as follows 
\bea
[e^{2A}\Im T]=v_A[\tilde\omega^A]+l_a[\tilde\chi^a]\in  \H^{1,1}_+\oplus \H^{2,2}_-\ .
\eea
Thus, in our description, $(\phi^a,t^A)$ are the chiral multiplets, while $(v_A,l_a)$ are the linear multiplets. Finally, in order to identify the conformal compensator $Y$ we have to fix a certain holomorphic $(3,0)$-form $\Omega^{0}_{\rm CY}$ and write
\bea
\Omega_{\rm CY}=Y^3\Omega^{0}_{\rm CY}\ .
\eea

The $\phi^a$-moduli clearly correspond to $\d$-closed shifts of the $B$ and $C_{\it 2}$ fields. If we write $[\Delta B]=b^a[\chi_a]$ and $[\Delta C_{\it 2}]=-c^a[\chi_a]$, then 
\bea
\phi^a=\frac{1}{g_s}\,b^a+ic^a\ .
\eea 
This deformation must be compensated by a corresponding deformation of $e^{-4A}J_{\rm CY}\wedge J_{\rm CY}$ in $\H^{2,2}_+$ of the kind discussed below, in order to guarantee that $[\hat\Delta\calt_{\it 4}]=0$. By construction, this deformation automatically satisfies the condition $\d_{H}[\Delta(e^{2A}\Im T)]=0$.

Before considering  the $h^{2,2}_+$  chiral fields corresponding to $\calt_{\it 4}$, it is convenient to first analyze the dual linear multiplets $v_A$ associated with $[e^{2A}\Im T]\in \H^{1,1}_+$, which can be clearly interpreted as the K\"ahler moduli of the underlying CY. On the other hand, one must impose (\ref{normCYcond}), obtaining
\bea\label{jomega}
v_A v_B v_C \,\cali^{ABC}=6|Y|^6\text{Vol}^{0}_{\rm CY}(M)\ ,
\eea
where
\bea
\cali^{ABC}:=\int_M\tilde\omega^A \wedge \tilde\omega^B \wedge\tilde\omega^C\quad,\quad  \text{Vol}^{0}_{\rm CY}(M):=-\frac{i}{8g_s}\int_M \Omega^{0}_{\rm CY}\wedge \bar\Omega^{0}_{\rm CY}\ .
\eea
From (\ref{jomega}) we see how the absolute value of the  conformal compensator can be seen as a function  of the CY K\"ahler moduli $v_A$.  This means that only $h^{1,1}_+-1$ of them give physically relevant deformations. Infinitesimally, they can be for example identified with the fluctuations $\delta v_A$ such that
\bea\label{primcoho}
v_A v_B\, \delta v_C \,\cali^{ABC}=0\ ,
\eea 
which correspond exactly to the primitive deformations. On the other hand, the universal K\"ahler structure deformation given by an overall rescaling of $J_{\rm CY}$ can be seen as a Weyl (i.e.\ pure gauge) transformation. This in turn implies that the linear multiplets can depend only on  $h^{1,1}_+-1$ of the $h^{2,2}_+=h^{1,1}_+$ chiral multiplets associated with $[\hat\Delta \calt]\in\H^{2,2}_+$. Let us now check this explicitly in the dual picture.

Indeed, at the infinitesimal level, the $\calt$ deformation dual to the unphysical linear multiplet corresponds to the infinitesimal deformation
\bea
\delta \Re T_{\it 4} =-\frac{1}2\, g_sr\,|Y|^{-4} J_{\rm CY}\wedge J_{\rm CY}\ ,
\eea   
which is generated by a shift $e^{-4A}\rightarrow e^{-4A}+r|Y|^{-4}$ (where the compensator appears to make $r$ of Weyl weight zero), which can always be integrated to a finite deformation since $e^{-4A}$ is completely determined by the supersymmetry conditions up to an arbitrary additional constant. From (\ref{wT}) it is clear that the corresponding deformation of $e^{2A}\Im T$ is a 6-form and thus it vanishes in $\d_H$-cohomology. So, the linear multiplets do not depend on the universal $\calt$-modulus 
\bea\label{univmodulus}
\rho=r+ic_r\ ,
\eea
where $c_r$ gives the RR-shift $[\Delta C]=(g_s/2)|Y|^{-4}c_r [J_{\rm CY}\wedge J_{\rm CY}]$, while they depend on the remaining $h^{1,1}_-+h^{2,2}_+-1$ $\calt$-moduli. However, extracting the explicit form of this dependence appears difficult since  the split of the $h^{2,2}_+$-moduli into universal and non-universal ones depends on  $J_{\rm CY}$. Furthermore, the $h^{2,2}_+-1$ non-universal deformations generically require compensating deformations of the $B$-field and the RR-fields.

Thus, it does not seem to be possible to extract the K\"ahler potential in general in an explicit closed form by following the procedure indicated in section \ref{sec:kaehler} and one has to study it case by case. This could be expected since already in the unwarped non-backreacted approximation in general an explicit form is not known \cite{louis}. However, as in that approximation, the general analysis simplifies drastically if we assume  $h^{1,1}_+=h^{2,2}_+=1$. In this case, we can take the generators of $\H^{1,1}_+$ and $\H^{2,2}_+$ to be
\bea
\tilde\omega=J^{0}_{\rm CY}\quad,\quad \omega=-\frac{J^{0}_{\rm CY}\wedge J^{0}_{\rm CY}}{6\text{Vol}^{0}_{\rm CY}(M)}\ ,
\eea  
where $J^{0}_{\rm CY}$ is the K\"ahler form that satisfies (\ref{normCYcond}) with respect to $\Omega^{0}_{\rm CY}$, so that $J_{\rm CY}=|Y|^2J^{0}_{\rm CY}$. 
Then, the unique CY K\"ahler modulus can be identified with the compensator, i.e.\ $v=|Y|^2$, and  we can write
\bea\label{uex}
[e^{2A}\Im T]=|Y|^2([\tilde\omega]+b^a[\tilde\omega\wedge\chi_a])\ ,
\eea
while
\bea
[\hat\Delta \Re T]=\frac{1}{g_s}b^a[\chi_a]+3g_sr\,\text{Vol}^{0}_{\rm CY}(M)[\omega]\ ,
\eea
with corresponding chiral fields $\phi^a$ and $\rho$ given by
\bea
[\hat\Delta \calt]=\phi^a[\chi_a]+3g_s\rho\,\text{Vol}^{0}_{\rm CY}(M)[\omega]\ .
\eea
Furthermore, at fixed $b^a$, we can identify the modulus $r$ by splitting
\bea\label{warpingsplit}
e^{-4A}=(e^{-4A^{0}}+r)/|Y|^4\ .
\eea
Using (\ref{uex}), we are led to identify the linear multiplets as follows
\bea
v=|Y|^2\quad,\quad l_a=|Y|^2\cali_{ab}b^b
\eea
where 
\bea
\cali_{ab}:=\int_M \chi_a\wedge \chi_b\wedge J^0_{\rm CY}\ .
\eea
The equations (\ref{linearchiral}) become in this case
\bea
v=\frac{1}{4\pi g_s \text{Vol}^{0}_{\rm CY}(M)}\frac{\del\caln_{\rm wCY}}{\del r}\quad,\quad l_a=\frac{3g_s}{4\pi}\frac{\del\caln_{\rm wCY}}{\del b^a}\ ,
\eea
which can be easily integrated to give
\bea
\caln_{\rm wCY}=\frac{4\pi}{3}|Y|^2\big[3g_s\text{Vol}^{0}_{\rm CY}(M)\,r+\frac1{2g_s}\cali_{ab}b^ab^b+\calc\big]\ ,
\eea
where $\calc$ is a constant. This constant can be determined by evaluating the complete expression (\ref{geomkaehler}) at $r=b^a=0$, obtaining
\bea
\calc=\frac{g_s}2\int_M e^{-4A^{0}} J_{\rm CY}^{0}\wedge J_{\rm CY}^{0}\wedge J_{\rm CY}^{0}\ .
\eea
Up to an additional constant, the resulting Einstein-frame K\"ahler potential is thus given by
\bea\label{wCYkaehler}
\calk_{\rm wCY}=-3\log\big[\rho+\bar\rho+\frac12\,\hat\cali_{ab}(\phi^a+\bar\phi^a)(\phi^b+\bar\phi^b)+\hat\calc\big]\ ,
\eea
with
\bea
\hat\calc=\frac{2\int_M e^{-4A^{0}} J_{\rm CY}^{0}\wedge J_{\rm CY}^{0}\wedge J_{\rm CY}^{0}}{\int_M  J_{\rm CY}^{0}\wedge J_{\rm CY}^{0}\wedge J_{\rm CY}^{0}}\quad,\quad  \hat\cali_{ab}:=\frac{\int_M \chi_a\wedge \chi_b\wedge J_{\rm CY}^{0}}{\int_M J_{\rm CY}^{0}\wedge J_{\rm CY}^{0}\wedge J_{\rm CY}^{0}}\ .
\eea  

\bigskip

By setting $\phi^a=0$ in (\ref{wCYkaehler}) one obtains the K\"ahler potential for the universal modulus found in \cite{torroba}.\footnote{See \cite{giddings,giddings2,douglas0,torroba0} for  previous related work, and \cite{fernandowarped} for a proposal, based on a probe D7-brane analysis, of warped K\"ahler potential including open string modes which reduces to the K\"ahler potential (\ref{wCYkaehler}) with $\phi^a=0$ once applied to the single universal closed string modulus.} The derivation presented in \cite{torroba} is based on a careful direct dimensional reduction in which supersymmetry does not play any particular role. On the contrary, here the interplay between 10D and 4D supersymmetry is crucial, allowing the above simple derivation of (\ref{wCYkaehler}) which keeps  also the $\phi^a$ moduli in the spectrum.  As in \cite{torroba}, in (\ref{wCYkaehler}) one can re-absorbe $\hat\calc$ in a shift of $r$, obtaining an expression which coincides with the unwarped one \cite{louis}. However, as stressed in \cite{torroba}, this shift could be physically non-innocuous and, for example, it can affect non-perturbative and $\alpha^\prime$-corrections.

Above we have identified the deformations $\hat\Delta \calt$ with representatives of $\H^{1,1}_-\oplus \H^{2,2}_+$. However, a key point is that generically one should consider them as classes of the $H$-twisted cohomology class $\H_H^{\rm ev}(M)_+$, via the isomorphism (\ref{wCYcc}). The same can be said about the use of $\H^{1,1}_+\oplus \H^{2,2}_-$ for $e^{2A}\Im T$, which should rather be considered as  
$\H_H^{\rm ev}(M;\mathbb{R})_-$. This means that for example
the basis  elements $\chi_a$ and $\omega_A$ of $\H_H^{\rm ev}(M;\mathbb{R})_+$ should be more generically considered as $\d_H$-closed and defined up to a $\d_H$-exact terms. (This implies that the generic representatives of $[\chi_a]$ and $[\omega_A]$ do not contain only  two-forms and four-forms, respectively.)  The relevance of this observation can be understood if for example one tries to write the instanton correction (\ref{instcorr}) (or the holomorphic coupling (\ref{holcoupling})) associated with a E3-brane (or D7-brane) wrapping a holomorphic four cycle $\Sigma$. Since $H\neq 0$, (in the untwisted picture) one must generically consider a non-vanishing $(1,1)$ and primitive gauge-invariant world-volume field-strength $\calf:=B|_\Sigma+{\rm F}$ (such that $\d\calf=H|_\Sigma$) \cite{fernandoland,lucal}. The instanton action (or D7-brane coupling) can be written as 
\bea
S_{\text{E3}}=S_{\text{E3}}^0+2\pi m_a \phi^a +2\pi n_A t^A
\eea
where, in the untwisted picture used in this section, the integrals
\bea
m_a:=\int_\Sigma \chi_a|_\Sigma\wedge e^\calf\quad,\quad n_A:=\int_\Sigma \omega_A|_\Sigma\wedge e^\calf 
\eea
can be considered as purely topological quantities (constant under generic deformations of $\Sigma$ and $\calf$), only if $\chi_a$ and $\omega_A$ are properly considered as classes of $\H_H^{\rm ev}(M;\mathbb{R})_+$, and not of $\H^{1,1}_-\oplus \H^{2,2}_+$. An analogous example regarding $\H_H^{\rm ev}(M;\mathbb{R})_-$ vs $\H^{1,1}_+\oplus \H^{2,2}_-$ is obtained by considering the coupling of D-string to the linear multiplets or the D-terms on space-filling D7-branes.

Let us briefly comment on the constant warping approximation. The pure spinors
(\ref{calzwCY}) and (\ref{wT}) reduce to the ones given in (\ref{purespinorCY}) if $g_sJ_{\rm CY}=e^{2A}J$ and $g_s\Omega_{\rm CY}=e^{3A}\Omega$, where $J$ and $\Omega$ describe the actual (non-rescaled) internal CY metric. Then, in this limit, the universal modulus (\ref{warpingsplit}) corresponds to a constant rescaling of the warping. This corresponds to a rescaling of the actual K\"ahler form $J$ (at fixed $J_{\rm CY}$) and thus coincides with the usual universal modulus. 

Finally, observe that the action of a E3-brane wrapping a divisor $\Sigma$ depends on the universal modulus $\rho$ as follows 
\bea
S_{\rm E3}=2\pi n\rho+\ldots\ ,
\eea where $n=-(1/2)g_s\int_\Sigma J^0_{\rm CY}\wedge J^0_{\rm CY}$, independently on the possible world-volume flux $\calf$. Thus, the corresponding non-pertubative superpotential involving the universal modulus does not depend on $\calf$ as well. The same is analogously true for the holomorphic coupling of space-filling D7-branes and thus  for the non-perturbative superpotential arising from gaugino condensation on a stack of them. As an example of a possible  consequence of this observation, several aspects of the KKLT proposal \cite{kklt} should not depend on the possible world-volume flux  $\calf$ on E3/D7 branes, which was assumed to be vanishing in that paper.\footnote{For example, assuming that this non-perturbative superpotential for the universal modulus can be used in presence of a supersymmetry-breaking flux $H^{0,3}\neq 0$, adding it to the expectation values $\langle\calw_{\rm GVW}\rangle$ of the GVW superpotential (\ref{gvw}) produces, together with the K\"ahler potential (\ref{wCYkaehler}) (with $\phi^a\equiv 0$), the same condition given in eq.~(13) of \cite{kklt}, which relates $\langle\calw_{\rm GVW}\rangle$ and $\langle\rho\rangle$. A similar result was obtained in section 6.1 of \cite{lucapaul2} by considering the 10D supersymmetry conditions modified by smeared E3/D7. A simple way to smear the E3/D7, proposed and discussed in detail in \cite{lucapaul2}, leads to a dependence of $\langle\calw_{\rm GVW}\rangle$ on $\langle\rho\rangle$ which, in presence of $\calf\neq 0$, differs  by a factor from the one given in eq.~(13) of \cite{kklt}. Thus, in order for the 10D approach of \cite{lucapaul2} to be in agreement the above 4D result, one should smear the E3/D7 in a different way. The possibility of different smearings was already suggested in \cite{lucapaul2}, although the details were not developed.}


\section{Other  subcases with SU(3)-structure}
\label{sec:othersu(3)}

One can consider other backgrounds with SU(3)-structure (see e.g.\ \cite{gianguido,freysu3,gmpt0}) and the simplest ones correspond to IIB backgrounds with O5-planes (and D5-branes) and IIA backgrounds with O6-planes (and D6-branes).  In these cases there is not an underlying CY or even K\"ahler metric surviving and thus it is difficult to describe microscopically the moduli and hence the K\"ahler potential. Furthermore, the tadpole condition generically requires the introduction of D-branes, which add new chiral fields mixing with the $\calt$ moduli and $U(1)$-gauge fields which can gauge the RR axial symmetry, producing D-terms as discussed in section \ref{Dtermmoduli}. Nevertheless we can still state what our general arguments predict about the $\calt$ spectrum ignoring these additional features.  These backgrounds may be thought of as flux and brane deformed CY manifolds and, as a  check, we will see that the spectrum coincides with the one obtained in the unwarped un-backreacted approximation \cite{louis}.  A more detailed study of these predictions is left for the future. 

\subsection{IIB SU(3)-structure vacua with O5-planes}

In this case\footnote{An extended discussion of vacua of this kind can be found in \cite{schulz}.}, taking into account orientifold projections (cf. appendix \ref{app:orientifold})
\bea
\calz=e^{3A-\Phi}\Omega\wedge e^B   \quad,\quad T=-ie^{-\Phi}e^{iJ+B}\ , 
\eea
where $(\Omega,J)$ defines an SU(3)-structure on the internal manifold. From the condition $\d\calz=0$  we get  a Calabi-Yau holomorphic structure 
\bea
\Omega_{\rm CY}=e^{3A-\Phi}\Omega\ .
\eea 
 On the other hand from $\d(e^{2A}\Im T)=0$ we get $e^{\Phi}=g_s e^{2A}$ with constant $g_s$, $H=0$ and
\bea\label{balanced}
\d(J\wedge J)=0\ .
\eea 
The remaining condition is 
\bea\label{sdd5}
\delbar(e^{-\Phi}J)=i F^{1,2}\ ,
\eea
which says that the space is actually not K\"ahler or, more precisely, does not have a K\"ahler structure naturally induced by the background supersymmetry.

The superpotential (\ref{supeff}) and its splitting into (\ref{supsplit}) now reads
\bea\label{d5sup}
\calw_{\rm eff}=\int_M\Omega_{\rm CY}\wedge F_{\it 3}=\int_M\Omega_{\rm CY}\wedge F^{\rm back}_{\it 3}+\int_{\Gamma_3}\Omega_{\rm CY}\ ,
\eea
where $\Gamma_3$ is a three-chain such that $\Sigma^{\rm D5}_2\subset \partial\Gamma_3$. The superpotential (\ref{d5sup}) can be studied along the lines described in \cite{D5sup}. 

The generalized Hodge decomposition coincides with the one given in (\ref{ghodgecomplex}) but, since the space is not required to be K\"ahler, the $\del\delbar$-lemma must be considered as an additional condition. However, thinking of these vacua as deformations of CY spaces induced by mutually supersymmetric O5-planes, D5-branes and fluxes, it is natural to consider the complex structure as unchanged by this deformation and the $\del\delbar$-lemma with it. Since $H=0$, this leads to the usual Hodge decomposition of cohomology and the condition (\ref{cohocond}) implies that
\bea
h^{3,2}=h^{1,0}=0\ ,
\eea
as for standard SU(3)-holonomy manifolds.

In this case, applying the orientifold projections of appendix \ref{app:orientifold}, our general arguments  predict the following spectrum of chiral fields 
\bea
[\hat\Delta \calt]\in \H^{1,1}_+\oplus \H^{2,2}_-\oplus \H^{3,3}_+\ .
\eea
On the other hand, the dual space of linear multiplets is given by
\bea
[e^{2A}\Im T]\in \H^{0,0}_+\oplus \H^{1,1}_-\oplus \H^{2,2}_+\ . 
\eea
Hence, we get exactly (part of) the massless field content  obtained in the unwarped approximation \cite{louis}, in which it corresponds to the NS deformations of K\"ahler structure, dilaton and $B$-field, completed  into chiral multiplets   by associated RR-moduli. However, let us stress again that in the backreacted picture these deformations could develop a different microscopical description (like for the universal modulus in the warped CY case) and could require to be accompanied by additional compensating deformations in order to solve the full coupled system of supersymmetry conditions, as for example it is evident from the fact that the internal space is not K\"ahler anymore.

\subsection{IIA SU(3)-structure vacua with O6-planes}

In this case, in terms of the SU(3)-structure $(\Omega,J)$ we have 
\bea
\calz=e^{3A-\Phi}e^{iJ+B}\quad,\quad T=e^{-\Phi}\Omega\wedge e^B\ .
\eea
Then, from $\d\calz=0$, one gets $H=0$, $e^{\Phi}=g_s e^{3A}$ and $\d J=0$, i.e.\ $J$ defines an ordinary symplectic structure. Indeed, the generalized complex structure defined by $\calj$ is just the $B$-transform of 
\bea
\calj=\left(\begin{array}{cc}  0 & J^{-1} \\  -J & 0
\end{array}\right)
\eea
which has the canonical form corresponding to a standard symplectic structure, in this case defined by $J$. The only non-vanishing RR-flux is $F_{\it 2}$ and the classical superpotential (\ref{supeff}) reduces to
\bea\label{f6d6}
\calw_{\rm eff}=\frac{1}{2g_s}\int_MF_{\it 2}\wedge J_{\rm c}\wedge J_{\rm c}\ ,
\eea
where $J_{\rm c}=J-iB$. Hidden in (\ref{f6d6}) there is also a contribution of the form (\ref{Dsup}) generated by D6-branes  \cite{lucasup},  which however is trivial at very low-energies since, even in the presence of fluxes, D6-branes wrap special Lagrangian cycles \cite{lucal}, which are classically unobstructed \cite{mclean,fernandotors}.

In order to compute the $\delbar$-cohomology, let us work in the untwisted picture, where the $B$-field appearing in $\calz$ is `rotated' away. Then, in this case, the $\d^\calj$ is given by
\bea
\d^\calj=[\Lambda,\d]=:\delta\ ,
\eea
where $\Lambda$ is the operator that contracts forms with the bivector $-J^{-1}$. 
One can show \cite{cavalcanti} that in this case the generalized Hodge decomposition (\ref{ghodge}) is isomorphic to the ordinary grading of forms through the isomorphism 
\bea
\varphi: \Lambda^{n-k}T^*_M\rightarrow U_k\quad,\quad \varphi(\omega_{\it k})= e^{iJ}e^{-\frac{i}{2}\Lambda}\omega_{\it k}\ .
\eea
Under this isomorphism the generalized Dolbeault operators $\bar\partial$ and $\del$ are just $\d$ and $\delta$:
\bea
\delbar\varphi(\alpha)=\varphi(\d\alpha)\quad,\quad \del\varphi(\alpha)=\varphi(\delta\alpha)\ .
\eea
Thus, the generalized Dolbeault cohomology classes are isomorphic to the standard de Rham cohomology
\bea\label{sympliso}
\H^k_{\delbar}(M)\simeq \H^{3-k}_{\rm dR}(M;\mathbb{C})\ .
\eea 
In this case,  the $\d\d^\calj$-lemma is equivalent to the so-called Lefschetz property (see e.g.\ \cite{cavalcanti}) and we see that the condition (\ref{cohocond}) is equivalent to 
\bea
\H^{1}_{\rm dR}(M;\mathbb{C})\simeq \H^{5}_{\rm dR}(M;\mathbb{C})=0\ .
\eea
 From (\ref{sympliso}) and the orientifold projections given in appendix \ref{app:orientifold}, we see that the spectrum of chiral multiplets is given by 
\bea
[\hat\Delta \calt]\in \H^3_{\rm dR}(M)_+\ ,
\eea
while the linear multiplets are given by
\bea
[e^{2A}\Im T]\in \H^3_{\rm dR}(M)_-\ .
\eea 
Again, we find agreement with the spectrum obtained for unbackreacted CY's with O6-planes and fluxes on top of them \cite{louis}, where $ \H^3_{\rm dR}(M)_+ $ corresponds to deformations of the dilaton and the CY complex-structure moduli, complexified by associated $C_{\it 3}$-moduli. Again, the microscopic description of these deformations generically changes once the backreaction is taken into account, since for example the fluxes break the integrability of the complex structure.


\section{Discussion and outlook}

This paper has suggested a new approach, based on the framework provided by generalized complex geometry, for investigating the low-energy effective theory describing type II warped flux compactifications to flat space.

 However, the results obtained represent only  a first step in this direction. Indeed, only the  $\calt$ closed string moduli have been explicitly included in the low-energy  effective theory, while open string moduli or other closed string moduli encoded in $\calz$ have not been explicitly considered.  Notice that,  if one can  guarantee a standard  effective low-energy description (for example by restricting to exactly flat $\calz$ and D-brane  moduli), supersymmetry imposes that the full K\"ahler potential should still be given by (\ref{geomkaehler}), where the D-brane dependence would enter essentially through equation (\ref{tbianchi}).  However, even in the simplifying assumptions of the present work, the resulting effective K\"ahler potential  is only implicitly defined and appears to generically depend on the microscopic details of the specific models. For this reason, it would be important to work out other examples besides the one discussed in section \ref{example:wCY}, where  explicit functional dependence of the K\"ahler potential can be extracted.
 
As discussed  in section  \ref{Dtermmoduli},  the inclusion of D-branes will generically generate D-terms for (some of) the $\calt$ moduli. It would be interesting to understand better in which regimes these D-terms can be directly included in the low-energy effective description.  A similar question arises if one considers the possibility of including all or part of the $\calz$ moduli that are lifted by the superpotential (\ref{supeff}), directly adding the superpotential itself to the effective action. Notice that, as for the K\"ahler potential (\ref{geomkaehler}),  the superpotential (\ref{supeff})  automatically includes the open string  superpotential (\ref{Dsup}). In any case,  the chiral fields describing D-brane deformations will generically  combine in a non-trivial way with both the $\calz$ and $\calt$ moduli and  a unifying fully coupled picture should consistently combine the results of \cite{lucapaul1} with the closed string picture suggested in this paper. Furthermore, the complete effective theory will also include vector multiplets, which have not been discussed here. Going beyond the classical level, as briefly mentioned in section \ref{probes}  the formalism developed  seems to naturally allow the inclusion of  non-perturbative effects arising from Euclidean D-branes. It would be interesting to see if it can also be helpful in the computation of the fermionic zero modes, along the lines of what happens for the bosonic zero-modes of space-filling D-branes \cite{lucapaul1}. I hope to come back to these points in future work.

The emerging physical picture rises a number a questions at the mathematical level as well. First of all, most of the derivations have been greatly simplified by assuming the $\d\d^\calj$-lemma (cf.~appendix \ref{lemma}), which is actually a property that could or could not be satisfied by a generalized complex manifold (counter-examples in which it is not satisfied could be provided by  compactifications on nilmanifolds\footnote{I thank Alessandro Tomasiello and Li-Sheng Tseng for remarks on this point.} \cite{tt,schulz,scan}). It would be interesting to see under which conditions the $\d\d^\calj$-lemma can be relaxed without substantially changing the results of this paper and, thus, preserving the encouraging self-consistency provided by  their physical interpretation. Other physically motivated assumptions that would require a better mathematical inspection are the condition (\ref{cohocond}) on the generalized Dolbeault cohomology of strictly minimally supersymmetric $N=1$ vacua and, more importantly, the non-degeneracy of the variational problems described in appendix \ref{fluxhitchin}. On top of these difficulties, the unavoidable inclusion of orientifolds complicates further a more complete understanding of the geometry of these vacua, which constitutes by itself a  challenging and still quite unexplored subject.


\vspace{1cm}

\centerline{\large\em Acknowledgments}

\vspace{0.5cm}

\noindent I would like to thank Alessandro Tomasiello for early collaboration on this project and many useful discussions and comments. Many thanks also to Michael Haack and Paul Koerber, for careful proofreading of the draft and many suggestions and remarks, and to Antoine Van Proeyen for clarifying discussions about superconformal supergravity. I would also like  to acknowledge Ralph Blumenhagen, Ilka Brunner, Gil Cavalcanti, Michael Douglas, Dieter L\"ust, Dimitrios Tsimpis and Angel Uranga  for useful discussions. This work  is supported by the DFG Cluster of Excellence ``Origin and Structure of  the Universe'' in M\"unchen, Germany.

\vspace{3cm}

\newpage

\centerline{\LARGE \bf Appendix}
\vspace{0.5cm}

\begin{appendix}

\section{Polyforms and $H$-twist in different pictures} 
\label{twist}

This paper adopts the paradigm of generalized geometry \cite{hitchin,gualtieri} which uses as fundamental objects polyforms
\bea
\omega=\sum_{k\text{ even/odd}}\omega_{\it k}
\eea 
 rather than differential forms of fixed degree. These polyforms can be seen as O(6,6) spinors, and the associated Clifford algebra can be identified with the generalized vectors $\mathbb{X}=X+\xi$, with $X\in T_M$ and $\xi\in T^*_M$, whose Clifford action is given by $\mathbb{X}\cdot\omega=\iota_X\omega+\xi\wedge \omega$. In particular even and odd polyforms can be seen as O(6,6) spinors of opposite chirality.
The use of polyforms gives the possibility to use an $H$-twisted differential $\d_H:=\d+H\wedge$ acting on them.  However, one can use different equivalent descriptions with corresponding different natural differentials ---  see \cite{hitchin2} for a related discussion in terms of gerbes. For the purposes of this paper,  we can distinguish between the following three main `pictures'. 

\bigskip

{\it Untwisted picture}

\noindent In this picture the generic polyform $\omega$ of definite parity
{\em does not transform} under the $B$-field gauge transformations $B\rightarrow B+\d\lambda$ and the natural differential is the $H$-twisted one 
\bea
\d_H\omega=\sum_{k\text{ even/odd}}(\d\omega_{\it k}+H\wedge \omega_{\it k-2})\ .
\eea
In this picture, at least in the way it arises in string theory, the physical information about the $B$-field (and its field-strength $H$) can be  encoded in the twisted differential $\d_H$. Furthermore, the generalized vector fields $\mathbb{X}$ are global sections of $T_M\oplus T_M^*$. Finally, the six-dimensional Hodge-$*$ operator   is defined as follows
\bea\label{hodge}
*(e^{a_1}\wedge\ldots \wedge e^{a_k})=\frac{1}{(6-k)!}\epsilon_{b_1\cdots b_{6-k}}{}^{a_k\ldots a_1}e^{b_1}\wedge\ldots \wedge e^{b_{6-k}}\ ,
\eea
where $e^{a}$ is a vielbein for the internal space $M$. Notice that this Hodge-$*$ does not  coincide with more usual ones because of a possible different degree-dependent overall sign. The definition (\ref{hodge}) is particularly convenient when dealing with polyforms since $*^2=-1$ independently on the degree of the form it is acting on.

\newpage

 {\it Twisted picture}
 
 \noindent In this picture the generic polyform $\omega$ of definite parity {\em transforms} under the $B$-field gauge transformation $B\rightarrow B+\d\lambda$ as follows
 \bea\label{bgaugeforms}
 \omega\rightarrow e^{\d\lambda}\wedge\omega\ .
 \eea
In this case, in the presence of a non-trivial $H$-field, a twisted polyform is represented on different patches  by ordinary polyforms that are related by transformations (\ref{bgaugeforms}). The natural differential is the ordinary exterior derivative $\d$, which indeed commutes with the gauge transformation (\ref{bgaugeforms}). 
In this picture the generalized vector fields $\mathbb{X}$ are sections of the extension bundle
\bea\label{extbundle}
0\rightarrow T_M^*\rightarrow E\rightarrow T_M\rightarrow 0\ .
\eea
They can be locally written as $\mathbb{X}=X+\xi$, with $X\in T_M$ and $\xi\in T^*_M$, and transform as $X+\xi\rightarrow X-\iota_X \d\lambda+\xi$ under (\ref{bgaugeforms}). The advantage of this picture is that it allows to encode the $B$-field degrees of freedom in twisted polyforms, as discussed  in appendix \ref{purespinors} for our physical setting.  One can go from the untwisted to the twisted picture (and vice-versa) by writing  $H=\d B$ locally  and then defining
\bea
\omega^{\rm twisted}=e^B\wedge\omega^{\rm untwisted}\ . 
\eea
Clearly $\d\omega^{\rm twisted}=e^B\wedge\d_H\omega^{\rm untwisted}$. In the twisted picture the natural Hodge operator is the $B$-twisted one
\bea\label{Bhodge}
*_{B}:=e^B*e^{-B}\ .
\eea 

\bigskip

{\it Mixed-twisted picture}

\noindent In this picture one fixes a certain reference flux $H_0$ in the cohomology class of $H$ and writes $H=H_0+\d B$ with $B$ now {\em globally} defined. In short, in this picture the polyforms are twisted with respect to the cohomologically trivial flux $\Delta H=\d B$ and the natural differential is $\d_{H_0}$.  

\bigskip

\noindent Most of this paper uses the twisted picture. The only exceptions are sections \ref{example:wCY} and \ref{sec:othersu(3)}, where the untwisted and half-twisted picture are also used. 

In all pictures there is a natural antisymmetric pairing, called the Mukai pairing, which associates with a pair of polyforms $\omega$ and $\chi$  the top-form
\bea\label{mukai}
\langle\omega,\chi\rangle:=[\omega\wedge\sigma(\chi)]_{\rm top}\ ,
\eea 
where the involution $\sigma$ acts as $\sigma(\omega_{\it k})=(-)^{k(k-1)/2}\omega_{\it k}$ on a ${\it k}$-form $\omega_{\it k}$. Since 
\bea
\langle e^B\wedge\omega,e^{B}\wedge \chi\rangle=\langle \omega,\chi\rangle\ ,
\eea the Mukai pairing is picture-independent.

Finally, in all three pictures the (un)twisted differentials define elliptic differential complexes with associated twisted cohomology groups which are all isomorphic and are denoted with $\H^\bullet_H(M)$ (writing  $\H^\bullet_H(M;\mathbb{R})$ if one restricts to real polyforms). We can split $\H^\bullet_H(M)$ as
\bea\label{twistedcoho}
\H^\bullet_H(M)=\H^{\rm ev}_H(M)\oplus\H^{\rm od}_H(M)\ ,
\eea
where, for later convenience, we identify $\H^{\rm ev}_H(M)$ with the cohomology group represented by odd polyforms in IIA and even polyforms in IIB, while $\H^{\rm od}_H(M)$ is represented by even polyforms in IIA and odd polyforms in IIB.
If $[\omega]$ and $[\chi]$ are classes in $\H^{\rm ev/odd}_H(M)$, then the integral 
\bea\label{intpairing}
\int_M\langle\omega,\chi\rangle
\eea
gives a well-defined non-degenerate antisymmetric pairing for $\H^{\rm ev/odd}_H(M)$, thus providing a generalized Poincar\'e duality.


\section{O(6,6) pure spinors and integrable generalized complex structure} 
\label{purespinors} 
 
Supersymmetric type II flux compactifications with SU(3)$\times$SU(3) structure can be characterized in terms of a pair of complex polyforms $\calz$ and $T$ that are  O(6,6) pure spinors \cite{gmpt}.\footnote{See \cite{lucapaul2} for more details about the definitions used here, up to renaming $t$ there with $T$ here and going to the twisted picture, and the appendix of \cite{nonsusy} for additional background material about conventions and notation.} Using the twisted picture, $\calz$ and $T$ encode the complete information about the NS sector as well as the internal spinors $\eta_1$ and $\eta_2$ defining the SU(3)$\times$SU(3) structure. By using the Clifford isomorphism associating bi-spinors to polyforms, $\calz$ and $T$  can be explicitly identified as follows: 
\bea\label{defps}
e^{-B}\wedge\calz\simeq -\frac{8i}{|a|^2}e^{3A-\Phi}\eta_1\otimes \eta_2^T\quad,\quad e^{-B}\wedge T\simeq -\frac{8i}{|a|^2}e^{-\Phi}\eta_1\otimes \eta_2^\dagger\ , 
\eea 
where $|a|^2:=||\eta_1||^2=||\eta_2||^2$. Using $\calz$ and $T$ as the fundamental variables, the metric (and thus the volume form $\d {\rm Vol}_6$) and the $B$-field do not depend on an overall point-dependent re-scaling of $\calz$ and $T$, whose normalizations determine the dilaton and warping through
\bea\label{dilatonwarp}
e^{-2\Phi}=\frac{\langle T,\bar T\rangle}{\d {\rm Vol}_6}\quad,\quad e^{6A}=\frac{\langle\calz,\bar\calz\rangle}{\langle T,\bar T\rangle}\ .
\eea
Furthermore, let us recall that the independent degrees of freedom contained in a pure spinor can be identified with its real (or equivalently imaginary) part, which must be a `stable' polyforms \cite{hitchin}. By this result, it follows that the complete NS plus spin-structure information is contained in $\calz$ and $\Re T$.\footnote{One could equivalently use the stable form $\Re \calz$ instead of $\calz$, but the latter automatically gives the natural complex complex structure on its moduli-space.} We recall that the compatibility condition defining the $SU(3)\times SU(3)$-structure can be written as
\bea\label{compa}
\langle \calz, \mathbb{X}\cdot \Re T\rangle=0 \quad\quad\forall \mathbb{X}\in E\ ,
\eea
that is indeed identically satisfied by (\ref{defps}).

The pure spinor $\calz$ defines a generalized  almost complex structure $\calj:E\rightarrow E$ whose  $+i$-eigenspace $L_\calj$ annihilates $\calz$, i.e.\ $L_\calj\cdot \calz=0$.\footnote{The same can be done by using the pure spinor $T$ which defines a different generalized almost complex structure, which however results not integrable because of the RR-fluxes and thus not directly relevant for this paper.} $\calj$ can be used to define the following decomposition of the space of polyforms \cite{gualtieri}
\bea\label{ghodge}
\Lambda^\bullet T^*_M\otimes \mathbb{C}=\sum_{k=-3}^3 U_{k}\quad\text{with}\ U_{k}:= \bar L_\calj^{3-k}\cdot\calz\ ,
\eea 
where $U_k=\overline{U_{-k}}$ and $\calz\in \Gamma(U_3)$. The $U_k$ space can be alternatively defined as the $ik$-eigenspace of $\calj$, which naturally acts  on polyforms  (see e.g.\ \cite{toma}). This characterization may be used to give an alternative definition of this action of $\calj$ on polyforms. Using the decomposition (\ref{ghodge}), the compatibility condition (\ref{compa}) can be rewritten as follows
\bea\label{compa2}
\Re T \in U_0\ .
\eea
 
The supersymmetry condition (\ref{susy1}) says that $\calz$ defines an integrable generalized Calabi-Yau structure \cite{hitchin}. This in turns implies that the generalized complex structure $\calj$ is integrable. Roughly, this means that it locally defines hybrid complex-symplectic coordinates \cite{gualtieri}. Thus, $\calj$ can define  as limiting cases ordinary  symplectic  (in IIA) or complex (in IIB) structures. For example, this  happens when the supersymmetry has ordinary SU(3)-structure, i.e.\ $\eta_1\propto \eta^*_2$ in IIA and $\eta_1\propto \eta_2$ in IIB.\footnote{One can find the explicit expression of the pure spinors $\calz$ and $T$ for the SU(3)-structure case in appendix A.3 of \cite{nonsusy}, taking into account that $\calz=e^{3A-\Phi}e^{B}\wedge\Psi_2$ and $T=e^{-\Phi}e^B\wedge \Psi_1$, where $\Psi_{1,2}$ are the untwisted pure-spinors used in that paper.}
 
The integrability of $\calj$ can be equivalently characterized by the requirement that the ordinary differential $\d$ (acting on twisted polyforms) splits as \cite{gualtieri}
\bea
\d=\partial+\bar\partial\ ,
\eea
where $\partial:\Gamma(U_k)\rightarrow \Gamma(U_{k+1})$ and $\bar\partial:\Gamma(U_k)\rightarrow \Gamma(U_{k-1})$.\footnote{In the untwisted picture we write $\d_H=\del_H+\delbar_H$.}

\section{Generalized Dolbeault cohomology and $\d\d^\calj$(or $\partial\bar\partial$)-lemma}
 \label{lemma}
  
 The operator $\delbar$ can be seen as a generalized Dolbeault operator that defines an elliptic complex, with associated cohomology
\bea\label{gendolb}
\H^\bullet_{\delbar}(M)=\bigoplus^3_{k=-3}\H^k_{\delbar}(M)\ .
\eea

In order to relate the generalized Dolbeault cohomology (\ref{gendolb}) and the twisted cohomology (\ref{twistedcoho}), one needs to assume that the generalized complex manifold satisfies the so-called $\d\d^\calj$-lemma (which is actually a property). Let us first introduce the following real differential
\bea\label{dj}
\d^\calj:=-i(\del-\delbar)\ .
\eea 
Notice that we can equivalently write $\d^\calj=[\d,\calj]$, where $\calj$ must be considered as an operator acting on polyforms, as discussed in appendix \ref{purespinors}. Then, the $\d\d^\calj$-lemma is satisfied if 
\bea\label{forlemma1}
\ker \d\cap {\rm Im}\,\d^\calj= \ker \d^\calj\cap {\rm Im}\,\d={\rm Im}\,\d\d^\calj\ .
\eea
Working with complex polyforms, one can equivalently say that a generalized complex manifold satisfies the (generalized) $\del\delbar$-lemma if 
\bea\label{forlemma2}
\ker \delbar\cap {\rm Im}\,\del= \ker \del\cap {\rm Im}\,\delbar={\rm Im}\,\del\delbar\ .
\eea
 
Assuming (as always  in this paper) that the $\d\d^\calj$-lemma is valid, it is possible to show \cite{cavalcanti} that the twisted cohomology  $\H^\bullet_H(M)=\H^{\rm od}_H(M)\oplus\H^{\rm ev}_H(M)$ can split as follows
\bea\label{cohodec}
\H_H^{\rm od}(M)&\simeq& \H_H^3(M)\oplus \H_H^1(M)\oplus\H_H^{-1}(M)\oplus\H_H^{-3}(M)\ ,\cr
\H_H^{\rm ev}(M)&\simeq& \H_H^2(M)\oplus \H_H^0(M)\oplus\H_H^{-2}(M)\ ,
\eea
where $\H_H^k(M)$ can be defined as the cohomology classes in  $\H^\bullet_H(M)$ that can be represented by elements of $U_k$ in the decomposition (\ref{ghodge}). Furthermore, one can prove that 
\bea\label{cohoiso}
\H_H^k(M)\simeq \H_{\delbar}^k(M)\ ,
\eea 
and thus  (\ref{cohodec}) can be seen as a generalized Hodge decomposition of the twisted cohomology in generalized Dolbeault cohomologies. 

Finally, the pairing given  by (\ref{intpairing}) splits into well-defined pairings on $\H_H^k(M)\times \H_H^{l}(M)$, or equivalently  $\H_{\delbar}^k(M)\times \H_{\delbar}^{l}(M)$, which respect the isomorphism (\ref{cohoiso}) and are non-vanishing and non-degenerate only if $k=-l$.
 
 \section{The orientifold action}
\label{app:orientifold}

In order to get consistent compactifications, orientifolds are required. They are described by an involution $\calo$ which is the combination of a background involution $\iota:M\rightarrow M$,  a world-sheet parity $\Omega$ and possibly a  factor $(-)^{F_L}$, which is sometimes needed to ensure $\calo^2=\bbone$. Requiring the fields to be invariant under $\calo$, for O$(3+n)$-planes (with $n=0,\ldots,6$) one  obtains the conditions \cite{pauldimi,lucapaul2}\footnote{The NS bosonic fields satisfy the usual projector conditions $\iota^*g=g, \iota^*B=-B$ and $\iota^*\Phi=\Phi$.}
\bea\label{or1}
\iota^*\calz=(-)^{\frac{n(n+1)}{2}}\sigma (\calz)\quad,\quad \iota^*\calt=(-)^{\frac{n(n-1)}{2}}\sigma (\calt)\quad,\quad \iota^*F=(-)^{\frac{n(n+1)}{2}}\sigma (F)\ .
\eea
Furthermore, we have that $\Im T$ must satisfy the projection condition
\bea\label{or2}
\iota^*\Im T=-(-)^{\frac{n(n-1)}{2}}\sigma (\Im T)\ .
\eea
Consistency with (\ref{BI}) then requires that the total current associated with space-filling D-branes and orientifolds must satisfy the projection condition
\bea\label{or3}
\iota^* j=-(-)^{\frac{n(n-1)}{2}}\sigma(j)\ .
\eea

The generalized complex structure $\calj$ defined by $\calz$ satisfies the following projection condition
\bea
\iota^*\calj=\cali\calj\cali^{-1}\ ,
\eea
where $\cali$ maps a generalized vector $\mathbb{X}=X+\xi$ to $\cali(\mathbb{X})=X-\xi$. Notice that the O-plane generalized tangent bundle $T_{\text{O-plane}}$, as defined in \cite{gualtieri}, is given by
\bea\label{toplane}
T_{\text{O-plane}}=\{\mathbb{X}\in E|_{\text{O-plane}}: \iota^*\mathbb{X}=\cali\mathbb{X}\}\subset E\ ,
\eea
where, since we are at the O-plane locus, $\iota^*$ acts on $\mathbb{X}$ only as an algebraic operator. Clearly, if $\mathbb{X}\in T_{\text{O-plane}}$ then $\calj\cdot\mathbb{X}\in T_{\text{O-plane}}$, since $\iota^*(\calj\cdot \mathbb{X})=\iota^*\calj\cdot \iota^*\mathbb{X}=\cali(\calj\cdot\mathbb{X})$, and so O-planes are generalized complex submanifolds, i.e.\ they solve the condition (\ref{gcsources}). On the other hand, (\ref{braneDterm}) is implied by the projection condition. Thus  O-planes wrap calibrated cycles by construction, as was already shown in  \cite{pauldimi} by spinorial methods.

We can split the twisted cohomologies in even and odd parts under the orientifold involution as follows. Let us first split the spaces entering the generalized Hodge decomposition as
\bea
\Gamma(U_k)=\Gamma_+(U_k)\oplus \Gamma_-(U_k)
\eea
with $\omega^+_k\in\Gamma_+(U_k)$ and $\omega^-_k\in \Gamma_-(U_k)$ satisfying
\bea
\iota^*\omega^\pm_k&=&\pm(-)^{\frac{n(n+1)}{2}}\sigma (\omega^\pm_k)\quad\quad\text{for $k$ odd}\ ,\cr \iota^*\omega^\pm_k&=&\pm(-)^{\frac{n(n-1)}{2}}\sigma (\omega^\pm_k) \quad\quad\text{for $k$ even}\ .
\eea
We can then write (\ref{or1}), (\ref{or2}) and (\ref{or3}) as
\bea\label{polyformsproj}
\calz,F\in \Gamma_+(U_{\rm odd})\quad,\quad \calt\in \Gamma_+(U_{\rm even})\quad,\quad \Im T, j\in \Gamma_-(U_{\rm even})\ ,
\eea
where $\Gamma_\pm(U_{\rm odd}):=\Gamma_\pm(U_3)\oplus \Gamma_\pm(U_1)\oplus\Gamma_\pm(U_{-1})\oplus\Gamma_\pm(U_{-3})$ and $\Gamma_\pm(U_{\rm even}):=\Gamma_\pm(U_2)\oplus \Gamma_\pm(U_0)\oplus\Gamma_\pm(U_{-2})$.  Going to cohomology, we then define
\bea\label{orcoho}
\H^{\rm od}_H(M)_\pm&=&\Gamma^{\text{closed}}_\pm(U_{\rm odd})/\d \Gamma_\pm(U_{\rm even})\ ,\cr  
\H^{\rm ev}_H(M)_\pm&=&\Gamma_\pm^{\text{closed}}(U_{\rm ev})/\d \Gamma_\mp(U_{\rm odd})\ ,\cr  
\H^k_{\delbar}(M)_\pm&=&\Gamma_\pm^{\text{$\delbar$-closed}}(U_k)/\delbar \Gamma_{\pm}(U_{k+1})\quad\quad \text{for $k$ odd}\ ,\cr
\H^k_{\delbar}(M)_\pm&=&\Gamma_\pm^{\text{$\delbar$-closed}}(U_k)/\delbar \Gamma_{\mp}(U_{k+1})\quad\quad \text{for $k$ even}\ .
\eea
As in (\ref{cohodec}), we can split  $\H^{\rm od}_H(M)_\pm$ and $\H^{\rm ev}_H(M)_\pm$ as
\bea
\H_H^{\rm od}(M)_\pm&\simeq& \H_H^3(M)_\pm\oplus \H_H^1(M)_\pm\oplus\H_H^{-1}(M)_\pm\oplus\H_H^{-3}(M)_\pm\ ,\cr
\H_H^{\rm ev}(M)_\pm&\simeq& \H_H^2(M)_\pm\oplus \H_H^0(M)_\pm\oplus\H_H^{-2}(M)_\pm\ ,
\eea
where $ \H_H^k(M)_\pm\simeq \H^k_{\delbar}(M)_\pm$. 

From (\ref{or1}), (\ref{or2}) and (\ref{or3}) we see that in presence of orientifolds one has the following projection conditions on the main cohomology classes discussed in this paper 
\bea\label{projrules}
[\calz]\in \H^{\rm od}_H(M)_+\quad,\quad [\hat\Delta\calt]\in \H^{\rm ev}_H(M)_+ \quad,\quad [e^{2A}\Im T]\in \H^{\rm ev}_H(M)_-\ .
\eea

Notice that the integral (\ref{intpairing}) on the orientifold covering space is not vanishing  only when the two polyforms belong to $\Gamma_\pm(U_{\rm odd})\times \Gamma_\pm(U_{\rm odd})$ or $\Gamma_\pm(U_{\rm even})\times \Gamma_\mp(U_{\rm even})$. Then (\ref{intpairing}) defines  non-degenerate pairings on $\H_H^k(M)_\pm\times \H_H^{-k}(M)_\pm$ for $k$ odd and $\H_H^k(M)_\pm\times \H_H^{-k}(M)_\mp$ for $k$ even, while in the other cases it vanishes.

Finally, the space of sections $\Gamma(E)$ of the extension bundle $E$ defined in (\ref{extbundle}) splits into $\Gamma_+(E)\oplus \Gamma_-(E)$, where
\bea
\Gamma_{\pm}(E)=\{\mathbb{X}\in\Gamma(E): \iota^*\mathbb{X}=\pm\cali\mathbb{X}\}\ .
\eea
In particular, the generalized diffeomorphisms (\ref{gendiff}) in presence of the orientifolded space are generated by sections of $\Gamma_+(E)$. From (\ref{toplane}), this implies that the sections of   $\Gamma_+(E)$ are tangent (in a generalized sense) to the orientifolds and thus the associated generalized diffeomorphisms leave the O-planes untouched (see \cite{lucasup,lucapaul1} for a description of the deformations of D-branes in the language of generalized complex geometry). Furthermore, the generalized diffeomorphisms generated by  $\Gamma_+(E)$ are compatible with the orientifolded cohomology groups  (\ref{orcoho}), in the sense that they do not change the corresponding cohomology classes.

\section{Hitchin-like functionals}
\label{fluxhitchin} 

First consider the supersymmetry condition (\ref{dflat}). This can be obtained as a D-flatness condition from the conformal K\"ahler potential $\caln$ defined in (\ref{geomkaehler}) \cite{lucapaul2}. In our language, this means the following. Fix a certain $\Re T^0$ that satisfies the F-flatness conditions (\ref{susy2bis}). Then consider the functional $\caln$ evaluated on the orbit generated by symmetry (\ref{Tgauge}), i.e.\ take  $\Re T=\Re T^0+(\d\Lambda)_0$  for generic real $\Lambda\in\Gamma(U_1\oplus U_{-1})$ and consider $\caln=\caln(\Lambda)$. From (\ref{1var}) one can easily see that $\caln(\Lambda)$ is extremized exactly at a point $\Lambda$ where (\ref{dflat}) is satisfied. The existence of a unique $\Lambda$ extremizing $\caln(\Lambda)$, up to the residual symmetry group (\ref{resym}), requires a non-degeneracy condition analogous to the one discussed by Hitchin in \cite{hitchin}, which he called $\d\d^J$-lemma (which does not coincide with the $\d\d^\calj$-lemma of appendix \ref{lemma}, but is actually implied by the latter \cite{toma}). More concretely, the Hessian is
\bea\label{hessianorient}
\delta^2\caln(\alpha_1,\alpha_2)=\frac{4\pi}{3}\int_Me^{2A}\left(\langle \d\alpha_1,J\cdot \d\alpha_2\rangle-\frac{4}{3}\frac{\langle \d\alpha_1,\Im T\rangle\langle \d\alpha_2, \Im T\rangle}{\langle\Re T,\Im T\rangle}\right)
\eea
where $J$ is the complex structure for polyforms introduced by Hitchin in \cite{hitchin}, which can be defined as follows. Use the generalized  {\em almost} complex structure defined by $T$ (which together with $\calj$ defines a generalized {\em almost} K\"ahler structure) to expand $U_0$ into $U_{0,3}\oplus U_{0,1}\oplus U_{0,-1}\oplus U_{0,-3}$ \cite{gualtieri}. Then $J$ takes value $-i$ on $U_{0,3}\oplus U_{0,1}$ and $i$ on  $U_{0,-1}\oplus U_{0,-3}$.    The Hessian (\ref{hessianorient}) is required to be non degenerate, up to the residual symmetry group (\ref{resym}), i.e.\ it is required to vanish for any $\alpha_1$ only if $\d\alpha_2$ is generated by an infinitesimal symmmetry transformation (\ref{resym}). 
Although the direct mathematical proof of such non-degeneracy appears difficult  at the present time, its validity is strongly supported by the requirement of having a consistent supersymmetric four-dimensional effective theory. Indeed,  as discussed in this paper, for the latter to appear sensible the moduli encoded in $\Re T$ should be counted exactly by $\H^{\rm ev}_H(M;\mathbb{R})$.

Viceversa, following \cite{toma}\footnote{See also \cite{witt} for a similar discussion in the unwarped approximation.} an analogous argument exists to argue that one can identify $e^{2A}\Im T$ with its cohomology class $[e^{2A}\Im T]$ in  $\H_H^{\rm ev}(M;\mathbb{R})$, since its representative is fixed by (\ref{tbianchi}), which derives from (\ref{susy2bis}). Let us define the pure spinor $\rho=-ie^{2A} T$, fix a cohomology class $[\Re\rho]$ in $\H_H^{\rm ev}(M;\mathbb{R})\simeq \H_H^0 (M;\mathbb{R})$ and thus write the generic representative of $[\Re\rho]$ in $U_0$ as $\Re\rho=\Re\rho_0+\d\d^\calj\alpha$, with real $\alpha\in U_0$. 
Then we can consider the functional 
\bea\label{funct3}
\calh(\alpha)=\frac{i}{8}\int_M\frac{\langle \rho,\bar\rho\rangle^2}{\langle\calz,\bar\calz \rangle}-\int_M\langle\alpha, j\rangle\ . 
\eea 
Under a general variation of $\delta\alpha$, we have
\bea
\delta\calh(\alpha)=-\int_M\langle\delta\alpha, \d\d^\calj(e^{-2A}\Im \rho) +j\rangle=\int_M\langle\delta\alpha, \d\d^\calj \Re T-j\rangle\ ,
\eea
showing that the functional (\ref{funct3}) is extremized exactly for $\alpha$ such that (\ref{tbianchi}) is satisfied.
The Hessian is now given by 
\bea
\delta^2\calh(\alpha_1.\alpha_2)=\int_Me^{-2A}\left(\langle\d\d^\calj\alpha_1,J\cdot{\d\d^\calj}\alpha_2\rangle +2\frac{\langle\d\d^\calj\alpha_1,\Im\rho\rangle\langle\d\d^\calj\alpha_1,\Im\rho\rangle}{\langle\Re\rho,\Im\rho\rangle} \right)
\eea
 As above, the non-degeneracy of the Hessian up to the symmetry group (\ref{resym}) appears difficult to prove. Nevertheless, the existence of a consistent  4D effective theory seems to indirectly require such non-degeneracy, since the moduli encoded in $\Re\rho$ should correspond to the scalar component of 4D linear multiplets and should be identified with $\H_H^{\rm ev}(M;\mathbb{R})$.

Finally, as in the bulk of the paper, these arguments can be extended to explicitly keep into account the presence of orientifolds as described in appendix \ref{app:orientifold}, by taking for example $\Lambda\in \Gamma_-(U_1\oplus U_{-1})$ and $\alpha\in \Gamma_+(U_0)$.

\end{appendix}





\end{document}